\documentclass[10pt,twocolumn,letterpaper]{article}

\usepackage{cvpr}              
\definecolor{cvprblue}{rgb}{0.21,0.49,0.74}
\usepackage[pagebackref,breaklinks,colorlinks,allcolors=cvprblue]{hyperref}
\usepackage{graphicx}
\usepackage{multirow}
\usepackage{multicol}
\usepackage{float}
\PassOptionsToPackage{numbers, compress}{natbib}
\usepackage{adjustbox}
\usepackage{array}
\usepackage{pifont}
\usepackage{stfloats}

\definecolor{annotation}{RGB}{0, 153, 0}
\definecolor{key_words}{RGB}{236, 0, 141}
\definecolor{RowColor}{rgb}{0.95, 0.95, 1}
\definecolor{cgray}{RGB}{220,220,220}
\definecolor{lightblue}{RGB}{163,199,235}
\definecolor{darkblue}{RGB}{0,76,153}
\definecolor{citegrey}{HTML}{75878a}
\definecolor{updatagreen}{RGB}{80,100,40}
\definecolor{updatagrey}{HTML}{686461}
\definecolor{citecolor}{HTML}{2980b9}
\definecolor{linkcolor}{HTML}{c0392b}
\makeatletter
  \newcommand\figcaption{\def\@captype{figure}\caption}
  \newcommand\tabcaption{\def\@captype{table}\caption}
\makeatother
\newcommand{\cmark}{\textcolor{black}{\ding{51}}}

\newcommand{\parahead}[1]{\vspace{0.7mm}\noindent\textbf{#1}}

\newcommand{\ourmodel}{Ctrl-RS\xspace}

\def\paperID{36622} 
\def\confName{CVPR}
\def\confYear{2026}

\title{Controllable Radar Simulation with Waveform Parameter Embedding}

\author{
\scalebox{0.95}{
\begin{minipage}{\textwidth}
\centering
{\normalsize
\textbf{Weiqing Xiao}$^{1,7}$\thanks{These authors contributed equally.} \quad
\textbf{Hao Huang}$^{2,7}$\footnotemark[1] \quad
\textbf{Chonghao Zhong}$^{3,7}$\footnotemark[1] \quad
\textbf{Yujie Lin}$^{2,7}$ \quad
\textbf{Nan Wang}$^{7}$ \quad
\textbf{Xiaoxue Chen}$^{4,7}$ \\
\textbf{Zhaoxi Chen}$^{5}$ \quad
\textbf{Saining Zhang}$^{5}$ \quad
\textbf{Shuocheng Yang}$^{6}$ \quad
\textbf{Pierre Merriaux}$^{8}$ \quad
\textbf{Lei Lei}$^{8}$ \quad
\textbf{Hao Zhao}$^{4,7}$\thanks{Corresponding author.} \\
\vspace{0.5mm}
$^1$NJU \quad
$^2$BJTU \quad
$^3$BIT \quad
$^4$AIR, THU \quad
$^5$NTU \quad
$^6$SVM, THU \quad
$^7$Lightwheel AI \quad
$^8$LeddarTech \\
\vspace{0.3mm}
{\tt\scriptsize weiqing001@smail.nju.edu.cn \quad zhaohao@air.tsinghua.edu.cn}
}
\end{minipage}
}
}

\begin{document}

\twocolumn[{%
\renewcommand\twocolumn[1][]{#1}%
\maketitle
\vspace{-3mm}
\centering
\includegraphics[width=\linewidth]{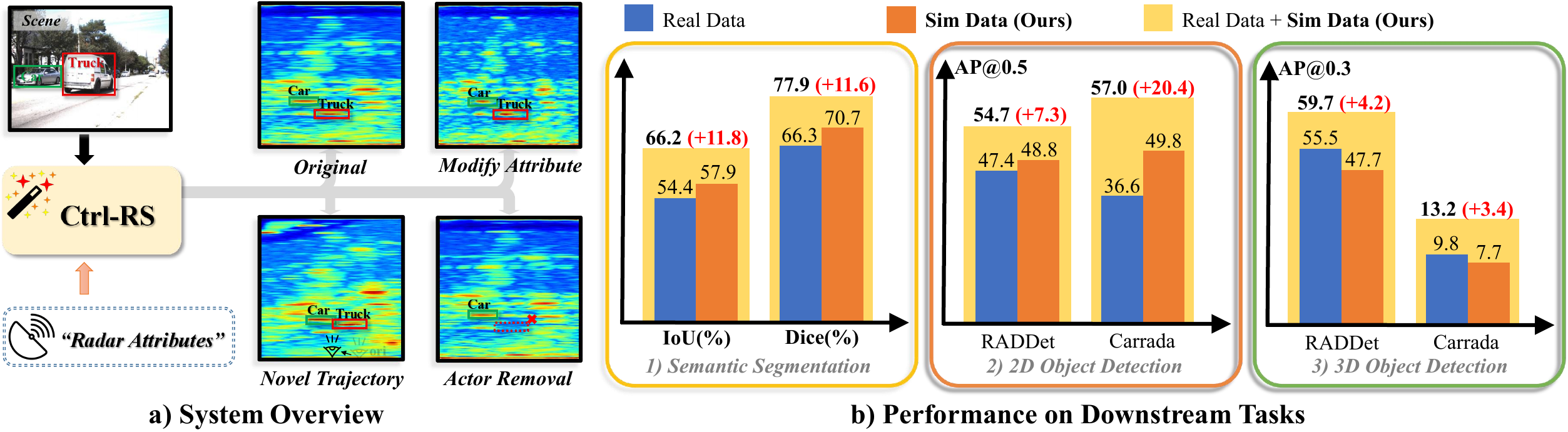}
\vspace{-6mm}
\captionof{figure}{
(a) \ourmodel{} enables controllable and realistic radar simulation by conditioning on customizable radar attributes. It supports flexible scene editing such as attribute modification, actor removal, and novel trajectories.
(b) \ourmodel{} improves performance on various tasks including semantic segmentation, 2D/3D object detection. In all settings, \ourmodel{}’s synthetic data either matches or surpasses real data, and provides consistent gains when combined with real-world datasets.
\vspace{1em}
}
\label{fig:radartop}
}]


\vspace{2mm}
\begin{abstract}
Autonomous driving simulators still lack high-fidelity radar, even though radar is critical for robust perception in adverse weather. A key obstacle is that raw radar point clouds are extremely sparse and stochastic, making it difficult to model; we argue that simulating the full range–azimuth–Doppler cube is a more principled target. Existing radar cube simulators either rely purely on neural generators, which are opaque and offer little control over sensor attributes, or on detailed electromagnetic pipelines, which are slow, require proprietary hardware specifications, and still struggle to capture real-world complexity. We introduce \textbf{\ourmodel{}}, a controllable radar cube simulation framework that combines the strengths of both worlds. First, we build an environment reflection tensor from diverse sensor sources (including LiDAR, monocular cameras, and existing radar). Second, we abstract radar physics into a compact set of waveform parameters that characterize the 3D point spread function, yielding an intuitive embedding of radar attributes such as range resolution, Doppler broadening, and azimuth beam shape. Third, we train a WARP-Net on a large mixed dataset that fuses real, analytically synthesized, and simulator-generated radar cubes to cover a wide distribution of radar attributes. \ourmodel{} supports viewpoint changes, actor removal, and attribute editing. Experiments on RADDet, Carrada, and nuScenes show that our simulated data can match or surpass real radar in 2D detection and semantic segmentation, and consistently boosts performance in 3D detection when combined with real data. The Project is available at \url{https://github.com/zhuxing0/Ctrl-RS}.
\end{abstract}

\vspace{-5mm}

\section{Introduction}
\label{sec:intro}

With its exceptional anti-interference capabilities and reliability under adverse weather conditions, radar plays a crucial role in advanced driver assistance systems (ADAS)~\cite{yan2024street,patole2017automotive, wu2025road,li2024uniscene,wang2025tacodepth,chu2025racformer,bang2025rctdistill}. However, its research and development have long faced challenges in data acquisition~\cite{zhou2024drivinggaussian,tonderski2024neurad,kung2025radarsplat,zhong2025cvfusion}. Consequently, the importance of radar simulation has become increasingly prominent. Radar simulation not only replicates diverse environmental conditions and traffic scenarios but also enables researchers to explore different radar parameter settings to cope with the diversity and complexity of real-world applications.

Radar emits electromagnetic waves and processes the reflected echo signals from targets, generating a radar cube (typically in the form of a range-azimuth-Doppler tensor) that records the relative velocity (Doppler shift) and reflectivity at each position~\cite{soumekh1999synthetic, li2008mimo, engels2021automotive}. This provides a detailed visualization of the surrounding environment~\cite{wang2025s3e,wu2025tars,lainon,yang2024v2x}. Existing radar simulation methods~\cite{ram2024synthesis,capsoni1998physically, gubelli2013ray, he2022channel, auer2016raysar} can be broadly categorized into two types: generative radar simulation and physics-based radar simulation.

Generative radar simulation directly generates domain-specific radar cube slices~\cite{huang2025towards,chae2025doppler,wheeler2017deep,haitman2025doppdrive} through generative adversarial networks (GAN)~\cite{ram2024synthesis, de2020generating, wang2020l2r} or variational autoencoders (VAE)~\cite{fidelis2023generation, weston2021there, wheeler2017deep}. Experimental results demonstrate that the generated synthetic data closely approximates real data in simple 2D tasks. However, this data-driven approach faces two major challenges: (1) Significant domain gaps exist in data distribution across different radar systems, necessitating separate data collection for each radar. (2) The absense of a physical prior prevents the extrapolation of radar attributes, limiting data simulation to predefined radar settings.

In contrast, physics-based radar simulation ~\cite{capsoni1998physically, yee1966numerical, clemens2001discrete, jin2015finite, machida2019rapid} simulates radar cube by modeling the complete process of an electromagnetic wave from emission to reception, ensuring interpretability and controllability over radar attributes. However, this approach demands precise radar hardware specifications, complex signal processing algorithms, and substantial computational time. Fortunately, recent progress~\cite{bialer2024radsimreal} provides us with new insights, which characterize radar attributes by the standard 3D reflection signal (i.e., the reflection signal of a single reflection point, also called the point spread function or PSF), although it still needs to be measured individually for each radar.

In this paper, we present \ourmodel{}, a radar simulation method that can handle different radar attributes simultaneously (Fig~\ref{fig:radartop}), focusing on simulating the range-azimuth-Doppler tensor (i.e., the radar cube). We provide a complete radar simulation pipeline, carefully designed from environment simulation to radar cube simulation. Specifically, we explore and validate the feasibility of constructing radar reflection information from existing sensor data for environment simulation. Concurrently, we propose a Waveform-Aware Radar Prediction (WARP-Net) to achieve controllable radar cube simulation. In WARP-Net, the radar attributes are characterized by the waveform parameters of the standard 3D reflection signal, which is not only visually intuitive, but also makes it effective and easy for WARP-Net to capture the signal variations under different radar attributes. We propose a series of methods to construct a mixed dataset enriched with radar attributes to robustly train WARP-Net.

We not only demonstrate the accuracy of \ourmodel{} simulation results on different datasets, but also experiment with them in several downstream tasks (including semantic segmentation and 2D/3D detection). Extensive experimental results show that baseline models trained on our simulation data can easily achieve comparable or even better performance than models trained on real data. When combining \ourmodel{}'s simulation data with real-world datasets, all downstream tasks experimented in this paper achieve further performance improvements. Furthermore, our \ourmodel{} successfully achieves realistic and stable simulations under various sensor input conditions, as well as in novel viewpoints and edited scenarios, demonstrating its potential for generating new scenarios in autonomous driving applications.

\section{Related Works}
\label{related works}

\paragraph{Physical radar simulation}\hspace{-1.5mm} primarily obtains reliable simulation data by modeling each stage of electromagnetic wave propagation—from emitting and reflecting to receiving and processing~\cite{soumekh1999synthetic, li2008mimo, engels2021automotive}. For example, to simulate reflected echo signals, researchers have proposed a series of time-domain electromagnetic simulations \cite{capsoni1998physically, yee1966numerical, clemens2001discrete, jin2015finite, machida2019rapid}, which model the behavior of electromagnetic fields over time by directly solving Maxwell's equations. While these methods are physically accurate, they are constrained by high computational requirements and stability issues, making them impractical for real-time autonomous driving simulations. Alternatively, ray-tracing based radar simulations \cite{auer2016raysar, yun2015ray, hirsenkorn2017ray, schussler2021realistic, gubelli2013ray, he2022channel} approximate signal propagation by modeling the interaction of electromagnetic waves with the environment. However, these methods require highly detailed hardware parameters and signal processing algorithms for the radar, which increases the complexity and computational requirements. Recently, RadSimReal~\cite{bialer2024radsimreal} proposes to use standard 3D reflection signal (i.e., PSF) to construct a direct link from the environment to the radar cube, thus overcoming the reliance on radar-specific details and specific algorithms (see the Appendix \S~\ref{para: Introduction to Radsimreal} for a more detailed description). However, it still requires collecting data for each unique radar sensor and manually measuring the corresponding standard 3D reflectance signal. Our \ourmodel{} draws on this idea and proposes to use waveform parameters of standard 3D reflection signal in different dimensions to characterize the radar attributes, which in turn enables effective control of radar attributes. 

\parahead{Generative radar simulation} aims to eliminate the need for detailed physical modeling of radar phenomena through a data-driven approach~\cite{dong2020probabilistic, meyer2021graph, jung20254d}, thereby bridging the domain gap with real-world data. For example, a series of works \cite{wang2020l2r, de2020generating, weston2021there} employ GAN or VAE to generate realistic radar cube slices. Additionally, several studies \cite{lei2024sar, borts2024radar, huang2024dart} have adopted NeRF to model 3D environments and generate novel view  for radar. Despite these advances, radar signals are still significantly affected by physical factors such as radar sensor attributes and environmental conditions. Ignoring these factors may lead to non-corresponding results~\cite{yun2015ray, hirsenkorn2017ray, schussler2021realistic}, i.e., the generated radar data does not satisfy a specific distribution in the real data. In contrast, our \ourmodel{} is a hybrid framework that not only achieves accurate generation of the complete radar cube through a data-driven approach (i.e., WARP-Net in this paper), but also incorporates physical prior (i.e., waveform parameters of standard 3D reflection signal) to capture radar signal variations under different radar attributes.
\section{Method}
\label{Method}

\begin{figure*}[t]
  \centering
  \includegraphics[width=\textwidth]{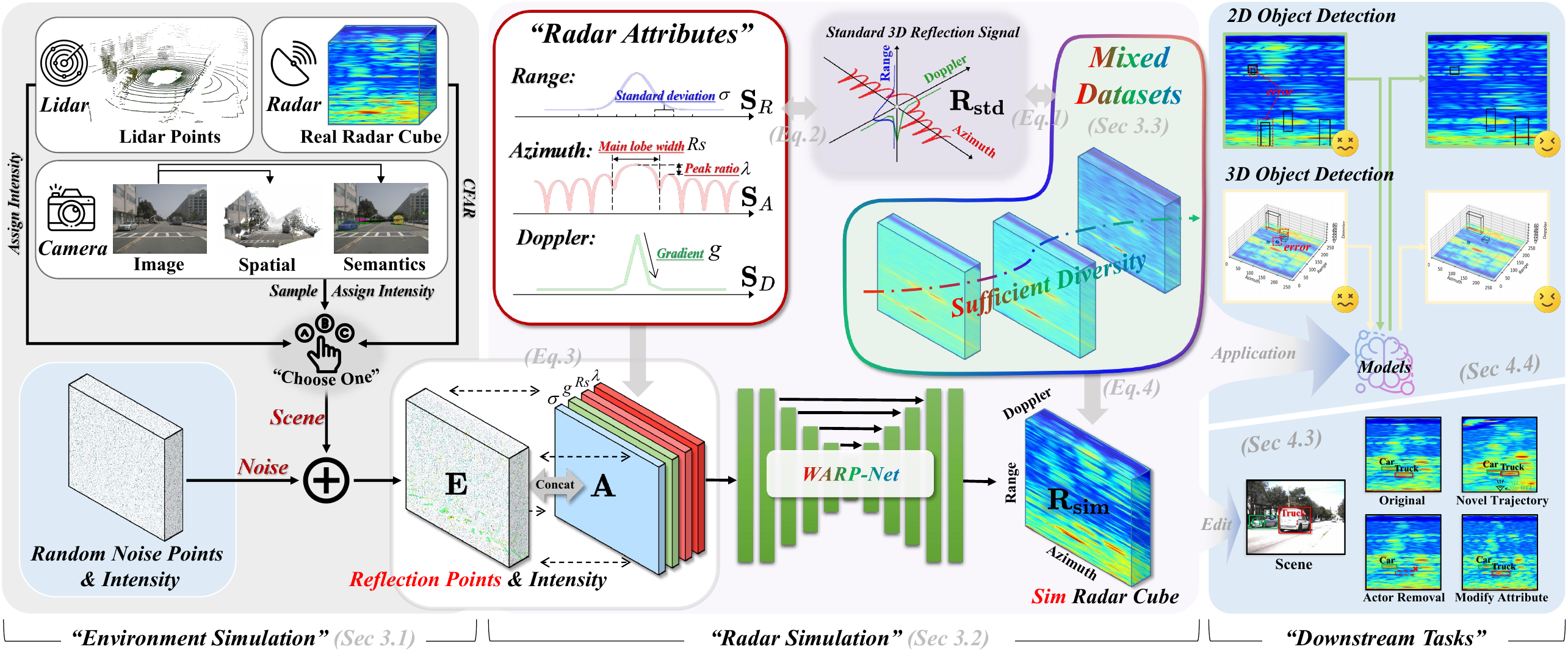}
  \vspace{-6mm}
  \caption{The overall framework of the proposed \ourmodel{}. The block diagrams from left to right show the environment simulation process, the radar simulation method and the downstream tasks of the simulation system in turn.}
   \label{fig:main}
   \vspace{-6mm}
\end{figure*}

As shown in Fig~\ref{fig:main}, our \ourmodel{} comprehensively designs the entire process from environment simulation (also known as environment construction) to radar simulation. In Sec~\ref{sec:es}, we first introduce environment simulation, a step traditionally regarded as a foundational module (independent of specific radar sensors) in prior methods. However, we explore a more flexible implementation approach that utilizes data from other sensors to achieve environment simulation. In Sec~\ref{sec:crs}, we present the core module of \ourmodel{}, a controllable radar simulation method (WARP-Net) capable of directly simulating the radar cube (i.e., range-azimuth-Doppler tensor) of the current environment based on configurable radar attributes. In Sec~\ref{sec:mdmt}, we describe how to construct a attribute-rich dataset for training WARP-Net.

\subsection{Environment Simulation}
\label{sec:es}

The environment simulation aims to obtain all the reflection points (i.e., the points that reflect the echo signals) and reflection intensities in the scene.
As shown in the left block diagram of Fig~\ref{fig:main}, in the environment simulation of \ourmodel{}, we construct a uniform reflection environment tensor $\mathbf{E}  \in \mathbb{R}^{r \times d \times a}$, where $r$, $d$, and $a$ denote the resolution of the radar cube in the Range-Doppler-Azimuth dimensions. In our approach, this tensor not only records the spatial locations, relative radar velocities, and reflection intensities of all reflective points in the current scene, but also considers noise signals within the radar cube (essentially false echoes such as thermal noise or multipath scattering).
Therefore, we divide the environment simulation into two parts: scene perception and noise simulation, and the specific details are described below.

\parahead{Scene Perception.} Unlike the traditional methods~\cite{schussler2021realistic,bialer2024radsimreal,luu2025rc,huang2025v2x,long2025riccardo,liang2025spatial} that strictly obtains reflection points through path tracing in graphics engines, we explore the feasibility of capturing scene reflection information through other sensors, such as LiDAR and camera. For example, we simply treat the LiDAR point cloud as reflected points and determine the reflection intensity of each point based on physical formulas~\cite{mahafza2003matlab}. Similarly, for camera image data, we use the reconstruction model VGGT~\cite{wang2025vggt} to obtain a dense scene point cloud. We then extract the target point cloud from it using instance masks generated by Grounded-SAM~\cite{ren2024grounded}. Finally, we obtain the reflected points and their reflection intensities through sparse random sampling and physical formulas~\cite{mahafza2003matlab}. Additionally, we support extracting reflection points and corresponding reflection intensities from existing radar data (using peak detection algorithms such as CFAR~\cite{farina1986review}). While reflection points obtained through these methods may lack precision, we experimentally verify \ourmodel{}'s robustness to them. This also implies the potential for directly supplementing radar information to a wide range of autonomous driving datasets.

\begin{figure*}[t]
  \centering
  \includegraphics[width=0.95\textwidth]{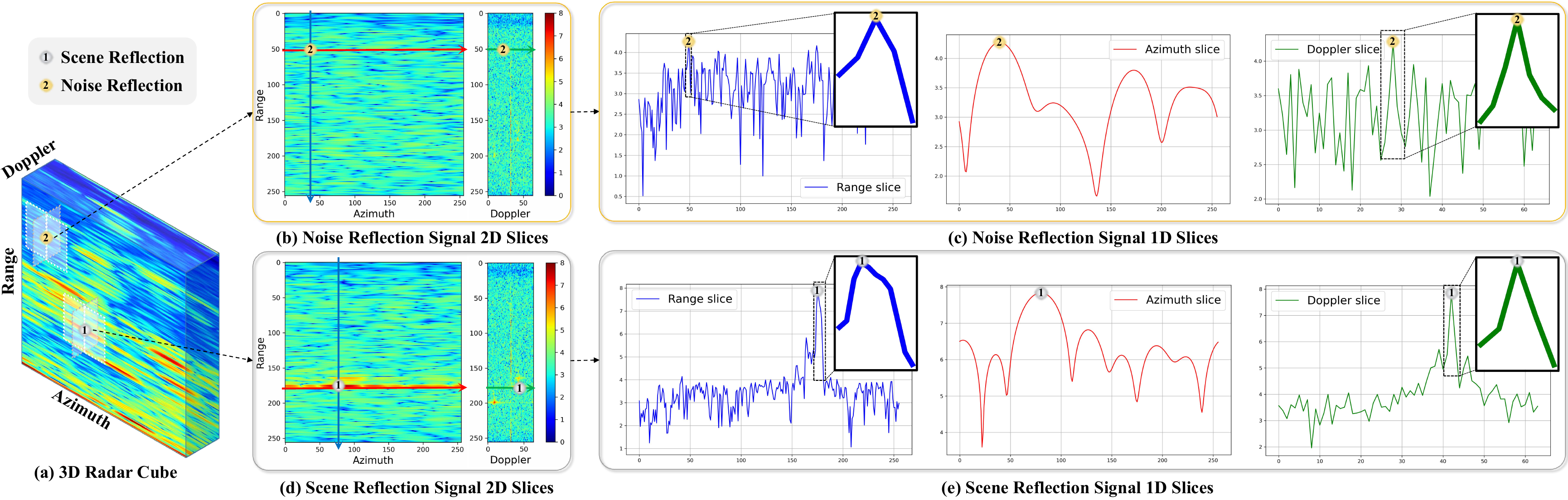}
  \vspace{-3mm}
  \caption{Visualization of 2D and 1D slices of radar cube from RADDet~\cite{zhang2021raddet}. In the 1D slices of the Range and Doppler dimensions, we box the range of signals reflected from the reflection points, outside of the box the reflected signals come from the other noise reflection points not labeled in the figure. 
  }
   \label{fig:radarslice}
   \vspace{-6mm} 
\end{figure*}

\parahead{Noise Simulation.} Our key observation is that the noise reflection signal in the real radar cube is very similar to the scene reflection signal in terms of waveforms and is randomly distributed in the radar cube, as shown in Fig~\ref{fig:radarslice}. Therefore, in order to more accurately reproduce the noise reflection signal in the radar cube, we propose to model the noise source using reflection points that are randomly distributed in all three dimensions: range, azimuth, and doppler (i.e., the noise signal is modeled as a radar signal from the noise reflection point, see the bottom left corner of Fig~\ref{fig:main}), instead of directly superimposing image noise (such as Gaussian noise) in the simulated radar cube as in the existing methods~\cite{bialer2024radsimreal,gubelli2013ray, he2022channel}.

After obtaining the reflection points and their reflection intensities from the scene and the noise, we map each reflection point to a corresponding position in the reflection environment tensor $\mathbf{E} $, and set the value of that position to the specific reflection intensity of that point.

\subsection{Controllable Radar Simulation}
\label{sec:crs}
The middle block diagram in Fig~\ref{fig:main} illustrates the radar simulation process of \ourmodel{}. The objective of this process is to simulate the radar cube based on the reflection environment tensor $\mathbf{E} $ described in Sec~\ref{sec:es} and to enable control over radar attributes. First, to overcome the need for radar-specific details in the simulation, we focus directly on the effect of different radars on the reflected waveform at each reflection point, and a waveform parameter-based representation of radar attributes (Sec~\ref{sec:rar}). Then, combining the advantages of both physical radar simulation methods and generative radar simulation methods, we propose a hybrid approach, WARP-Net (Sec~\ref{sec:arg}). On the one hand, it achieves efficient and realistic simulations through data-driven and implicit learning approaches. On the other hand, it effectively captures the variation of radar cubes under different radar attributes by integrating the reflection environment tensor $\mathbf{E} $ and waveform parameter-based radar attribute embeddings.

\subsubsection{Waveform Parameter-based Radar Attribute}
\label{sec:rar}

Following RadSimReal~\cite{bialer2024radsimreal}, we model the radar cube $\mathbf{R}  \in \mathbb{R}^{r \times d \times a}$ as a weighted superposition of standard 3D reflection signal $\mathbf{R_{\mathrm{std}}}$ (i.e., the PSF in RadSimReal~\cite{bialer2024radsimreal}) at each reflection point according to the reflection intensity:
\begin{equation}
\mathbf{R}  = \sum_{i=1}^{K} I_i \cdot \mathbf{R_{\mathrm{std}}}(r_i, d_i, a_i)
\end{equation}
where $I_i$ denotes the reflection intensity of the $i$-th reflection point, $(r_i,d_i,a_i)$ denote the corresponding coordinates of the $i$-th reflection point in the radar cube, and $K$ denotes the total number of reflection points. This modeling allows us to characterize radar attributes by the 3D waveform of $\mathbf{R_{\mathrm{std}}}$ without the need to obtain specific details of radar. Surprisingly, we observe that the 3D waveforms of $\mathbf{R_{\mathrm{std}}}$ are similar among different radar datasets~\cite{zhang2021raddet,zhang2023peakconv,wang2021rodnet} (see Fig~\ref{fig:mdataset1d} in Appendix). Inspired by this, we propose a waveform parameter-based radar attribute representation.

\begin{figure}[t!]
    \centering
    \includegraphics[width=0.48\textwidth]{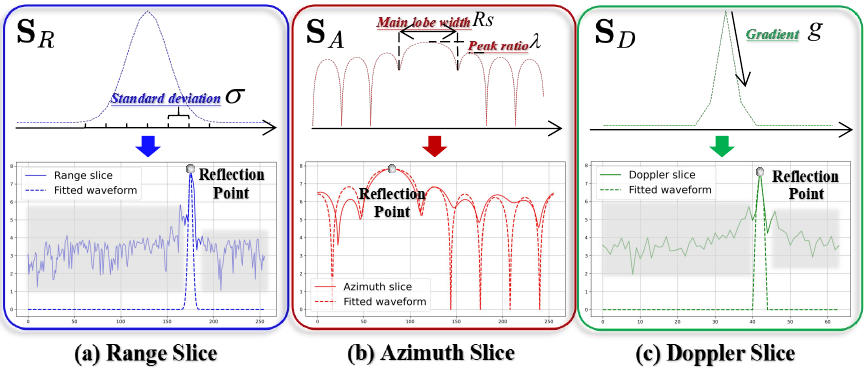}
    \vspace{-7mm}
    \caption{Visualization of the fitting functions and fitting results. Note that for the 1D slices, we only show the fitting results for a single reflection point, and the parts of the Range slice and Doppler slice that are covered with a transparent gray block actually belong to the other reflection points.}
    \label{fig:radarwave}
    \vspace{-0.6cm}
\end{figure}

Specifically, we first fit slices of this 3D waveform in different dimensions using different functions (see Appendix \S~\ref{para: Standard 3D reflection Signal on Different Datasets} for function formulas):
\begin{equation}
\mathbf{R_{\mathrm{std}}}(r_i,d_i,a_i)=\mathbf{S}_{R}(r_i)*\mathbf{S}_{D}(d_i)*\mathbf{S}_{A}(a_i)
\end{equation}
where $\mathbf{S}_{R}$ is the Gaussian function, capturing the Range resolution and decay. $\mathbf{S}_{D}$ is the segmented linear function, modeling Doppler broadening. $\mathbf{S}_{A}$ is the spectral function of the window function, describing the Azimuth beam pattern. We then characterize the radar attributes by the standard deviation $\mathbf{\sigma}$ of $\mathbf{S}_{R}$, the gradient $g$ of $\mathbf{S}_{D}$, and the main lobe width $Rs$ and peak ratio $\lambda$ in $\mathbf{S}_{A}$ (theoretically, more parameters could characterize radar attributes more accurately, but in practice, these parameters are sufficient for simulation and are easily obtainable through measurement methods~\cite{ludbrook1997comparing}). As shown in Fig.~\ref{fig:radarwave}, by adjusting the parameters ($\mathbf{\sigma}$, $g$, $Rs$, and $\lambda$), the proposed functions can accurately fit the 3D reflection signal $\mathbf{R_{\mathrm{std}}}$ at a single reflection point. 

\subsubsection{Controlling the Radar via Attribute Embedding}
\label{sec:arg}

Unlike RadSimReal~\cite{bialer2024radsimreal} which performs separate reflection waveform measurements for each type of radar, our goal is to train a radar simulation network with definable radar attributes. Thanks to our proposed waveform parameter-based representation of radar attributes (Sec~\ref{sec:rar}), we can define radar attributes with a few simple parameters. On this basis, we propose WARP-Net, which directly simulates the radar cube based on the reflection environment tensor $\mathbf{E} $ (Sec~\ref{sec:es}) and the waveform parameter-based radar attribute. Due to page constraints, we provide its architecture details in the Appendix \S~\ref{para: Network Structure of WARP-Net}. In summary, we first construct the radar attribute embedding $\mathbf{A} \in \mathbb{R} ^{4\times r\times d\times a}$, which is filled with $\mathbf{\sigma}$, $g$, $Rs$, $\lambda$ in order on its 4 channels. Then, we concatenate the radar attribute embedding $\mathbf{A} $ and the reflection environment tensor $\mathbf{E} $ , and process them with the WARP-Net: 
\begin{equation}
\mathbf{R_{\mathrm{sim}}} = U(\mathrm{Concat}\left\{ \mathbf{E} ,\ \mathbf{A}  \right\})
\end{equation}
where $\mathrm{Concat}$ denotes the Concat operation, and $U$ denotes the WARP-Net (the final output is processed using the Relu function), and $\mathbf{R_{\mathrm{sim}}}\in \mathbb{R} ^{r\times d\times a}$ is the simulation radar cube. 


\subsection{Mixed Datasets and Model Training}
\label{sec:mdmt}

In this section, we describe how to construct the training dataset (Sec~\ref{sec:md}) and the training strategies (Sec~\ref{sec:t}) for our WARP-Net. Full implementation details are provided in the Appendix.

\subsubsection{Mixed Datasets}
\label{sec:md}

For the proposed WARP-Net, both the diversity of radar attributes and the quality of radar cube in the training dataset are crucial. However, the existing real radar datasets only cover a limited number of radar attributes. To overcome this challenge, we design two efficient data augmentation methods, creating a mixed dataset rich in radar attributes based on existing real-world radar data. Depending on the collection route, the mixed radar dataset can be divided into 3 parts: the real dataset A, the synthetic dataset B, and the simulation dataset C. The motivation and ideas for constructing these three parts are described below (see the Appendix \S~\ref{para: Mixed Datasets} for specific details).

\parahead{Real dataset A.} It consists of multiple radar cube datasets collected from real-world scenarios~\cite{zhang2021raddet,ouaknine2021carrada}. For each radar cube dataset, we measure the waveform parameters of the radar cube through formula fitting, ultimately selecting the average value of the results as the radar attribute label.

\parahead{Synthetic dataset B.} We draw inspiration from RadSimReal~\cite{bialer2024radsimreal} to obtain the synthetic dataset B. Specifically, we generate multiple 3D reflection waveforms by presetting the dense waveform parameters, and then calculate the convolution of the reflection waveforms at each reflection point according to the method in RadSimReal, and finally obtain the synthetic radar cube. Overall, although synthetic dataset B is of lower quality than real dataset A, it covers a much richer set of radar attributes.

\parahead{Simulation dataset C.} To improve the overall quality of the mixed dataset, we generate the simulation dataset C using the original 3D U-Net (distinct from WARP-Net) without incorporating attribute embeddings. Specifically, we train a separate 3D U-Net on each real radar cube dataset, i.e., each model weight is dedicated to learning radar simulation under one radar attribute. Although not capable of capturing radar attributes, each trained 3D U-Net can generate accurately the radar cube for a specific radar attribute in different scenarios. We inference these 3D U-Nets on their respective unseen scenarios to generate a series of high-quality radar cubes. For example, assuming the existence of $N$ real datasets, these correspond to $N$ different radar attributes and $N$ scene sets. By pairing them in all possible combinations, we can generate $N*(N-1)$ new simulation datasets, significantly enriching our mixed dataset.

\subsubsection{Training}
\label{sec:t}
We use L1 loss to train our WARP-Net. However, due to the random and dense distribution of the noisy reflection signals in all dimensions of the radar cube (Fig~\ref{fig:radarslice}), the number of noisy reflection points in the reflection environment tensor $\mathbf{E} $ is much larger than the scene reflection points. To ensure the accuracy of the scene reflection signal, we additionally compute the loss at all scene reflection point locations and name it $L_{\mathrm{scene}}$. The total loss is as follows:
\begin{equation}
L = \left\| \mathbf{R_{\mathrm{sim}}} - \mathbf{R_{\mathrm{gt}}} \right\|_1 + L_{\mathrm{scene}}
\end{equation}
in which
\begin{equation}
L_{\mathrm{scene}} = \left\| \mathbf{R_{\mathrm{sim}}}[P_{s}] - \mathbf{R_{\mathrm{gt}}}[P_{s}] \right\|_1
\end{equation}
where $\mathbf{R_{\mathrm{gt}}}$ is the ground truth radar cube and $P_{s}$ denotes the coordinate index of all scene reflection points in the radar cube (and also in $\mathbf{E} $).

\section{Experiments}
\label{sec:experiments}


\subsection{Implementation Details}
We perform our experiments on a single NVIDIA A800 GPU. For WARP-Net, we train the model for 50 epochs with a batch size of 3, using the AdamW optimizer and a one-cycle learning rate schedule with a learning rate of $2\times 10^{-4}$. All experimental datasets are collected from real-world road scenarios, with $r=256$, $a=256$, and $d=64$. Training ($bs=3$) and inference ($bs=1$) require 20.5 GB and 2.5 GB of GPU memory, respectively. The training takes around one day. For a fair comparison, the models in each downstream experiment are trained using the same optimization strategy. See the Appendix for further details and additional experimental results, including the dataset description and ablation studies of training datasets, etc.

\begin{figure*}[t!]
    \centering
    
    \begin{minipage}[t]{0.49\textwidth}
        \centering
        \includegraphics[width=\textwidth]{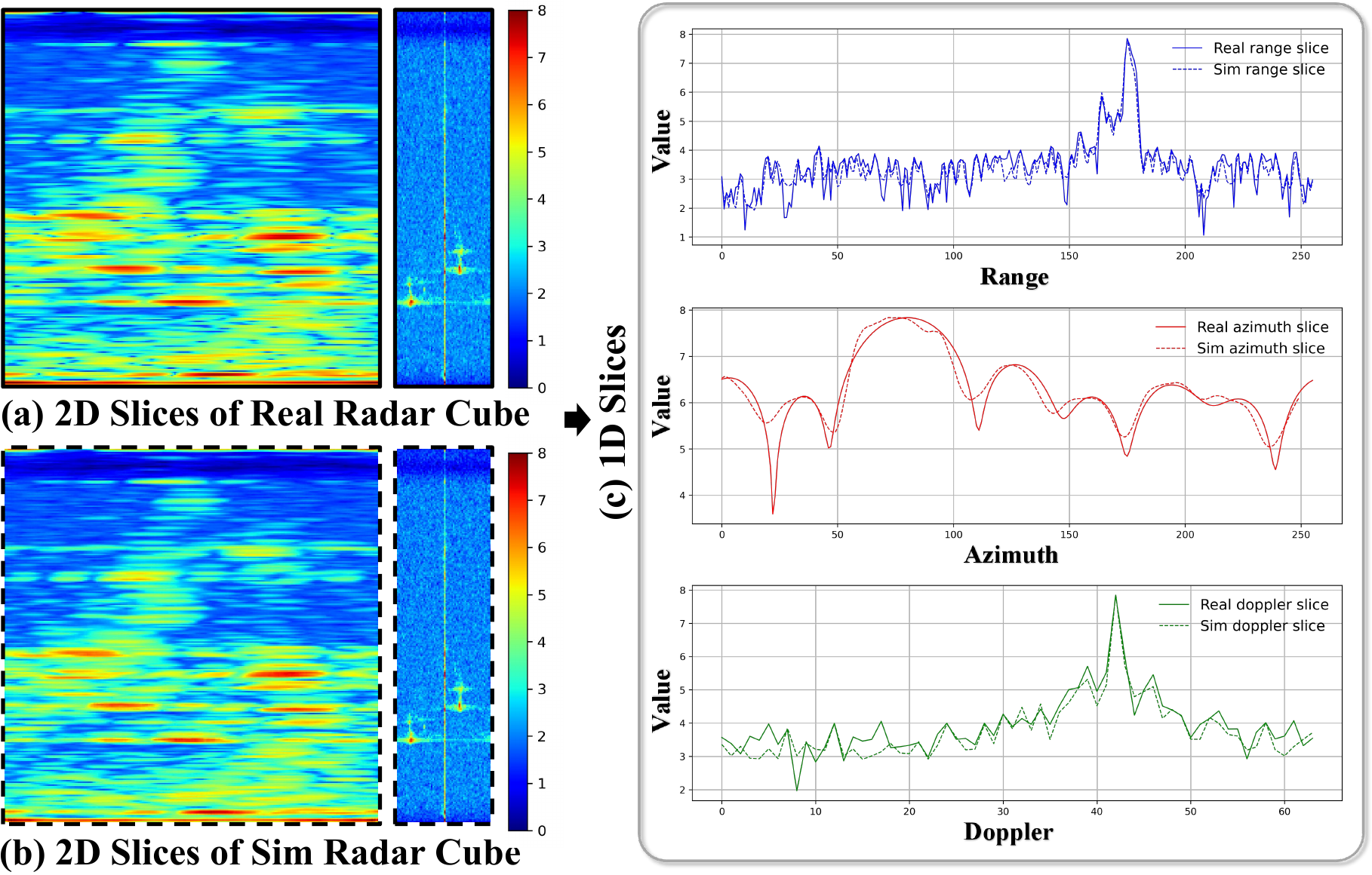}
        \vspace{-7mm}
        \figcaption{Comparison between the real radar cube and our simulation radar cube on RADDet~\cite{zhang2021raddet}. 
        }
        \label{fig:radarcompare}
    \end{minipage}
    \hfill
    \raisebox{22mm}{
    \begin{minipage}[t]{0.48\textwidth}
        \centering
        \includegraphics[width=0.9\textwidth]{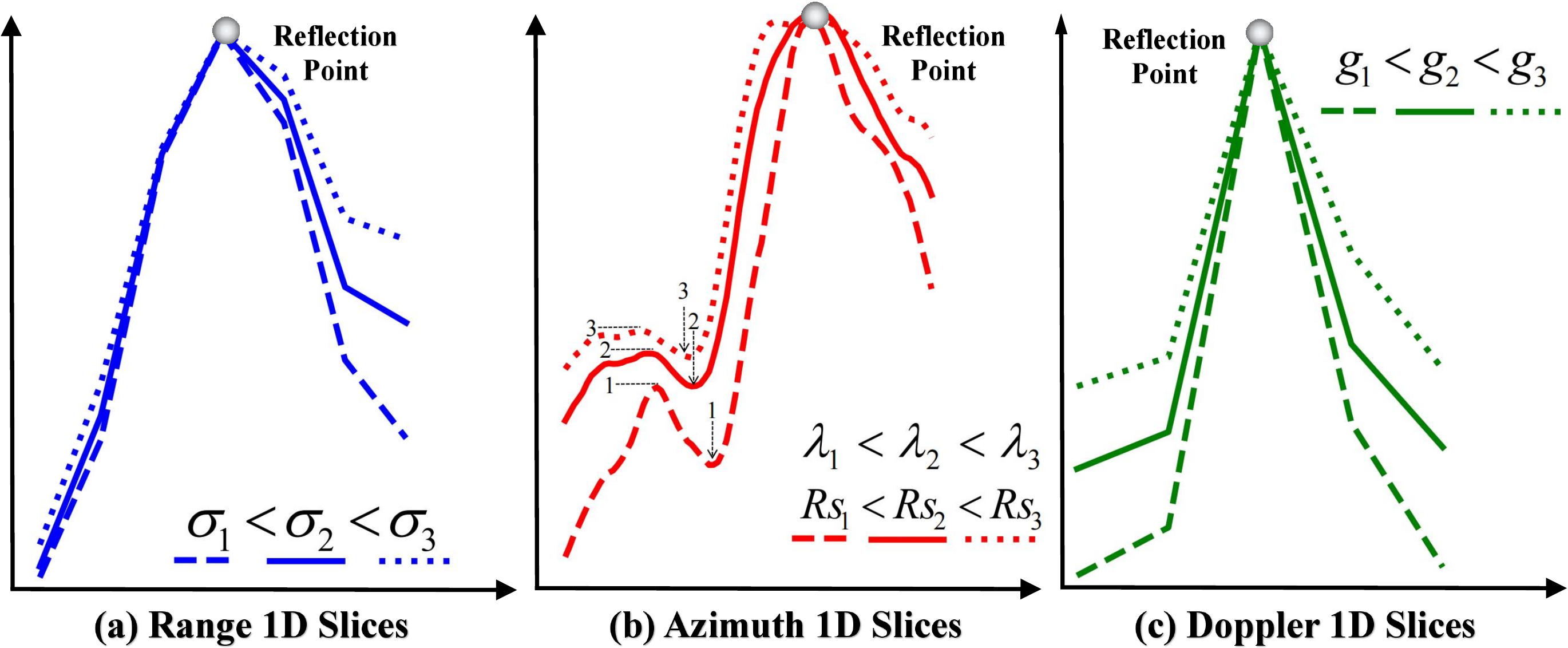}
        \vspace{-3mm}
        \figcaption{Comparison of 1D slices of the simulation results for the same reflection point with different waveform parameters. We intercept the part of the 1D slices that is affected by the labeled reflection point.}
        \label{fig:radarae}
    \end{minipage}}
    
    \vspace{-10mm}

    \begin{minipage}[t]{0.48\textwidth}
        \centering
        \footnotesize
        \setlength{\tabcolsep}{2pt}
        \renewcommand{\arraystretch}{0.95}
        \vspace{-1mm}
        \resizebox{\linewidth}{!}{
        \begin{tabular}{l|ccccc|ccccc}
            \toprule[1pt]
            \multirow{2}{*}{Method}  & \multicolumn{5}{c|}{RADDet} & \multicolumn{5}{c}{Carrada} \\
            \cmidrule(l){2-6} \cmidrule(r){7-11}
            & $\mathrm{PPE}$ & $\mathrm{PPE}_{s}$ & PPSE & FID & $\mathrm{T_{sim}}$ & $\mathrm{PPE}$ & $\mathrm{PPE}_{s}$ & PPSE & FID & $\mathrm{T_{sim}}$ \\
            \midrule[0.5pt]RadSimReal~\cite{bialer2024radsimreal} & 0.644 & 0.022 & 48.0 & 18.8 & 0.61s & 0.634 & 0.063 & 48.9 & 45.2 & 0.65s \\
            \midrule[0.5pt]
            WARP-Net w/o $L_{\mathrm{scene}}$ & \textbf{0.260} & \underline{0.023} & \textbf{19.9} & \underline{1.4} & \multirow{2}{*}{\textbf{0.04}s} & \textbf{0.261} & \underline{0.038} & \textbf{20.0} & \underline{20.0} & \multirow{2}{*}{\textbf{0.04}s} \\
            WARP-Net w/ $L_{\mathrm{scene}}$ & \underline{0.267} & \textbf{0.009} & \underline{20.0} & \textbf{1.1} & & \underline{0.271} & \textbf{0.012} & \underline{20.1} & \textbf{12.6} & \\
            \bottomrule[1pt]
        \end{tabular}}
        \vspace{-2mm}
        \tabcaption{Quantitative comparisons across different datasets. PPE and PPSE denote per-point error and per-point spectral error, respectively. The $s$ denote reflected point locations. Our WARP-Net is trained on the proposed mixed dataset (completely isolated from the test sets). See Appendix \S~\ref{para: Formula for Calculating} for calculation of above metrics.
        }
        \label{tab:radartimeandsim}
        \vspace{2.65mm}
    \end{minipage}
    \hfill
    \raisebox{12mm}{
    \begin{minipage}[t]{0.48\textwidth}
        \centering
        \footnotesize
        \setlength{\tabcolsep}{2pt}
        \renewcommand{\arraystretch}{0.92}
        \vspace{-3mm}
        \resizebox{0.95\linewidth}{!}{
            \begin{tabular}{l|l|c|c|c}
                \toprule[1pt]
                Test Set & \multicolumn{1}{c|}{Train Set} & AP & AP@0.5 & AP@0.75  \\
                \midrule[0.5pt]
                \multirow{5}{*}{RADDet} 
                & RADDet (R.) & 24.9 & 47.4 & 22.3 \\
                & Sim-R-by-R + Sim-R-by-C & 25.9 & 48.8 & 24.3  \\
                & R. + Sim-R-by-R & \underline{27.6} & 51.2 & 25.0 \\
                & R. + Sim-R-by-C & \underline{27.6} & \underline{52.2} & \underline{25.2} \\
                & R. + Sim-R-by-R + Sim-R-by-C & $\textbf{28.5}_{\textcolor{red}{\textbf{+3.6}}}$ & $\textbf{54.7}_{\textcolor{red}{\textbf{+7.3}}}$ & $\textbf{25.7}_{\textcolor{red}{\textbf{+3.4}}}$  \\
                \midrule[0.5pt]
                \midrule[0.5pt]
                Test Set & \multicolumn{1}{c|}{Train Set} & AP & AP@0.5 & AP@0.75  \\
                \midrule[0.5pt]
                \multirow{5}{*}{Carrada} 
                & Carrada (C.) & 12.8 & 36.6 & 6.4 \\
                & Sim-C-by-C + Sim-C-by-R & \underline{24.0} & 49.8 & \underline{19.6} \\
                & C. + Sim-C-by-C & 21.7 & 48.6 & 15.7 \\
                & C. + Sim-C-by-R & 22.7 & \underline{50.4} & 16.3 \\
                & C. + Sim-C-by-C + Sim-C-by-R & $\textbf{29.1}_{\textcolor{red}{\textbf{+16.3}}}$ & $\textbf{57.0}_{\textcolor{red}{\textbf{+20.4}}}$  & $\textbf{26.4}_{\textcolor{red}{\textbf{+20.0}}}$ \\
                \bottomrule[1pt]
            \end{tabular}
        }
        \vspace{-3mm}
        \tabcaption{2D Object detection performance (AP) on RADDet and Carrada using RTMDet-Tiny~\cite{lyu2022rtmdet} as baseline. See Fig~\ref{fig:radartop} for bar chart. \textcolor{red}{Red numbers} represent improvements over the baseline (same applies to following tables).}
        \label{tab:RC2D}
    \end{minipage}}
    \vspace{-6mm}

\end{figure*}

 \subsection{Analysis of Simulation Results}
To verify the utility of the proposed \ourmodel{}, we evaluate the system in various aspects, including simulation quality, simulation efficiency, and attribute controllability. The following are the specific evaluation results.

\parahead{Simulation Quality and Efficiency.}
To accurately assess the similarity between simulation and real data on the same scene, we extract all reflection points and reflection intensities (both scene and noise) from the real radar cube using the CFAR algorithm. We then simulate the radar cube using WARP-Net. Table~\ref{tab:radartimeandsim} records the simulation error and time consumption of both methods on different datasets. Our method achieves per-point errors below $0.27$ on both RADDet~\cite{zhang2021raddet} and Carrada~\cite{ouaknine2021carrada}. After introducing $L_{\mathrm{scene}}$, although the global error increases slightly, the error of the scene reflection signal decreases significantly, which proves the effectiveness of the supervision. In addition, the global error of our method is significantly smaller compared to RadSimReal~\cite{bialer2024radsimreal}, as we accurately model the widely distributed noise reflection signals in the radar cube. Fig~\ref{fig:radarcompare} shows the visualization results. Our simulation results are highly similar to the real radar cube, including the noise reflection signal.
In terms of efficiency, our radar simulation takes around 0.04s, which is much faster than RadSimReal and traditional physical simulations~\cite{auer2016raysar, yun2015ray, hirsenkorn2017ray, schussler2021realistic} (which generally take at least 5s). Additionally, we also record the full runtime of our method under different scene perception modes, as shown in Table~\ref{tab:nuscene_ra}. Overall, the majority of our method's computational cost stems from environmental simulation, yet it can still be recognized as an efficient radar simulation approach.


\parahead{Effectiveness of Attribute Embedding.}
In order to verify the ability of \ourmodel{} to capture different radar attributes, we compare the simulation results of the radar cube with different waveform parameters. As shown in Fig~\ref{fig:radarae}, our method can effectively sense the changes of radar attribute embedding and accurately align the waveforms in the radar cube. The simulation results under all radar attributes are geometrically similar to the real radar cube, which strongly demonstrates the physical reliability of our method.

\subsection{Novel Sensor Trajectory}
In this section, we explore \ourmodel{}'s ability to synthesize a radar cube in a new viewpoint. Fig~\ref{fig:radartop} (a) shows the results of removing the target in the scene and moving the viewpoint 5 meters laterally (see the Appendix \S~\ref{para: Details of Novel Sensor Trajectory} for more results and implementation details). Our \ourmodel{} successfully reconstructs realistic radar cubes on new viewpoints and edited scenes, demonstrating its potential for generating new scenes in autonomous driving applications.

\vspace{-2mm}

\subsection{Downstream Validation}

In this section, we explore the application value of \ourmodel{} simulation data through a variety of different downstream tasks. To exclude gains due to scenario diversity, we generate simulation data only on the training set scenarios of RADDet~\cite{zhang2021raddet} and Carrada~\cite{ouaknine2021carrada} (see Appendix \S~\ref{para: Datasets} for a detailed description of these two datasets). For the radar attribute embedding, we set two sets of waveform parameter sets $\left \{ \sigma, g, Rs, \lambda \right \} _{R}$ and $\left \{ \sigma, g, Rs, \lambda \right \} _{C}$, which cover but are not limited to the radar attributes of RADDet and Carrada, respectively. Then, we generate four simulation datasets: \textbf{1) Sim-R-by-R}, which is generated on the RADDet training set scenario with the radar attribute from $\left \{ \sigma, g, Rs, \lambda \right \} _{R}$. \textbf{2) Sim-R-by-C}, which is generated on the RADDet training set scenario with the radar attribute from $\left \{ \sigma, g, Rs, \lambda \right \} _{C}$. \textbf{3) Sim-C-by-C}, which is generated on the Carrada training set scenario with the radar attribute comes from $\left \{ \sigma, g, Rs, \lambda \right \} _{C}$. \textbf{4) Sim-C-by-R}, which is generated on the Carrada training set scenario with the radar attribute comes from $\left \{ \sigma, g, Rs, \lambda \right \} _{R}$.

In the following, we show the experiment results on each task specifically (more experimental details are provided in the Appendix, along with comparisons to RadSimReal~\cite{bialer2024radsimreal} and standard data augmentation). 

\begin{table}[t]
    \centering
    \setlength{\tabcolsep}{4pt}
    \setlength{\abovecaptionskip}{-2pt}
    \setlength{\belowcaptionskip}{2pt}
    \renewcommand{\arraystretch}{1.1}
    \footnotesize
    \begin{adjustbox}{max width=\linewidth}
    \begin{tabular}{l|cc|cc}
        \toprule[1pt]
         \multicolumn{1}{c|}{\multirow{2.5}{*}{Train Set}} & \multicolumn{2}{c|}{IoU (\%)} & \multicolumn{2}{c}{Dice (\%)}  \\
         \cmidrule(l){2-3} \cmidrule(r){4-5}
         & RA & RD & RA & RD  \\
        \midrule[0.5pt]
         Carrada (C.) & 39.2 & 54.4 & 48.5 & 66.3 \\
         Sim-C-by-C + Sim-C-by-R & 37.0 & 57.9 & 44.9 & 70.7 \\
         C. + Sim-C-by-C & \textbf{41.2} & \underline{65.1} & \textbf{51.2} & \underline{76.7} \\
         C. + Sim-C-by-R & 39.6 & 63.4 & 49.1 & 75.0 \\
         C. + Sim-C-by-C + Sim-C-by-R & \underline{40.2} & \textbf{66.2} & \underline{49.7} & \textbf{77.9} \\
        \bottomrule[1pt]
    \end{tabular}
    \end{adjustbox}
    \vspace{2mm}
    \caption{Performance comparison of multi-view semantic segmentation on the Carrada test set. The baseline is MVRSS~\cite{ouaknine2021multi}, which analyzes RA and RD views of continuously acquired radar signals to semantically segment them. See Fig~\ref{fig:radartop} for bar chart. Complete results for each category are provided in Table~\ref{tab:RCseg_all} from Appendix.}
    \vspace{-5mm}
    \label{tab:RCseg}
\end{table}

\parahead{2D Object Detection.} First, we evaluate the effectiveness of the simulation data on the 2D object detection task. Using RTMDet-Tiny~\cite{lyu2022rtmdet} as the baseline, we conduct five groups of experiments on the RADDet and Carrada datasets, under identical training settings for fair comparison. Objects are detected on Range-Doppler slices of the 3D radar cube, which are treated as 2D images. This differs from the 2D YOLO Head of RADDet, which performs 3D detection by outputting bounding boxes in the range-azimuth plane.
The quantitative results are presented in Table~\ref{tab:RC2D}. We observe that training solely on our simulation data already outperforms training on real data. Moreover, joint training with both simulation and real data achieves the best performance. These results demonstrate that \ourmodel{}'s simulation data can not only serve as a substitute for real data in 2D detection, but also enhance diversity and provide effective data augmentation.

\begin{table*}[t]
    \small
    \centering
    \vspace{2mm}
    \resizebox{2\columnwidth}{!}{
    \begin{tabular}{l|l|c|c|c|c|c|c}
        \toprule[1pt]
        \multirow{2}{*}{Test Set} & \multicolumn{1}{c|}{\multirow{2}{*}{Train Set}} & \multicolumn{6}{c}{\multirow{1}{*}{AP@0.3 of RAD / RA}}  \\
         \cmidrule(lr){3-8} 
         & & Person & Bicycle & Car & Bus & Truck & All \\
         \midrule[0.5pt]
         \multirow{5}{*}{RADDet} & RADDet (R.) & 32.91 / 51.80 & 22.25 / 39.88 & 65.42 / 81.20 & 43.42 / 42.11 & 51.55 / 62.63 & 55.50 / 70.91 \\
         & Sim-R-by-R + Sim-R-by-C & 26.05 / 54.50 & 7.80 / 46.34 & 56.71 / 78.00 & 36.84 / 47.37 & 46.88 / 67.15 & 47.72 / 70.48 \\
         & R. + Sim-R-by-R & 34.22 / 56.72 & 25.14 / \underline{57.66} & 67.94 / 83.29 & \underline{44.74} / \underline{50.00} & \underline{54.18} / \textbf{71.63} & 57.72 / 75.37 \\
         & R. + Sim-R-by-C & \underline{36.25} / \textbf{63.68} & \textbf{30.64} / 57.37 & \underline{68.00} / \underline{85.18} & \textbf{52.63} / 47.37 & 52.70 / 69.36 & \underline{58.25} / \underline{77.50} \\
         & R. + Sim-R-by-R + Sim-R-by-C & \textbf{36.85} / \underline{62.97} & \underline{27.75} / \textbf{61.13} & \textbf{69.25} / \textbf{85.94} & \underline{44.74} / \textbf{60.53} & \textbf{57.64} / \underline{70.38} & $\textbf{59.66}_{\textcolor{red}{\textbf{+4.16}}}$ / $\textbf{78.02}_{\textcolor{red}{\textbf{+7.11}}}$ \\
        \bottomrule[1pt]
    \end{tabular}
    }
    \vspace{-2mm}
    \caption{3D Object detection AP@0.3 on RADDet. The baseline is the RADDet model~\cite{zhang2021raddet}, which takes as input a 3D radar cube and predicts 2D boxes via a RA YOLO Head and 3D boxes via a RAD YOLO Head. See Fig~\ref{fig:radartop} for bar chart.}
    \vspace{-5mm}
    \label{tab:RC3Dallc_raddet}
\end{table*}

\begin{table}[t]
    \small
    \centering
    \vspace{2mm}
    \resizebox{\columnwidth}{!}{
    \begin{tabular}{l|l|cc}
        \toprule[1pt]
        \multirow{1}{*}{Test Set} & \multicolumn{1}{c|}{\multirow{1}{*}{Train Set}} & RAD & RA \\
        \midrule[0.5pt]
        \multirow{5}{*}{Carrada} & Carrada (C.) & 9.83 & 29.91 \\
        & Sim-C-by-C + Sim-C-by-R & 7.68 & 29.19 \\
        & C. + Sim-C-by-C & 12.10 & \underline{31.39} \\
        & C. + Sim-C-by-R & \underline{13.05} & 28.25 \\
        & C. + Sim-C-by-C + Sim-C-by-R & $\textbf{13.21}_{\textcolor{red}{\textbf{+3.38}}}$ & $\textbf{31.55}_{\textcolor{red}{\textbf{+1.64}}}$ \\
        \bottomrule[1pt]
    \end{tabular}
    }
    \vspace{-2mm}
    \caption{3D Object detection AP@0.3 on Carrada. The baseline is the RADDet model~\cite{zhang2021raddet}, which takes as input a 3D radar cube and predicts 2D boxes via a RA YOLO Head and 3D boxes via a RAD YOLO Head. See Fig~\ref{fig:radartop} for bar chart. Complete quantitative results for each category are provided in Table~\ref{tab:RC3Dallc_all} from Appendix.}
    \label{tab:RC3Dallc_carrada}
\end{table}

\parahead{Multi-View Radar Semantic Segmentation.}
In order to verify the semantic accuracy of the simulation data of \ourmodel{}, we conduct experiments on MVRSS~\cite{ouaknine2021multi}. MVRSS is a multi-view semantic segmentation method based on the radar cube, which analyzes RA (Range-Azimuth), RD (Range-Doppler) views of consecutive frames to semantically segment them. Same as other downstream tasks, we train MVRSS on different combinations of datasets and then directly compare their performance on a real test set. Table~\ref{tab:RCseg} illustrates the quantitative comparison results. The model trained entirely on our simulation data achieves better performance than the model trained on the real dataset. In addition, we achieve further improvements when using joint training with simulation data and real data. These results demonstrate not only the semantic accuracy of these simulation data, but also the stability of \ourmodel{}'s simulation on dynamic scenes.


\begin{table}[t]
    \centering
    \begin{adjustbox}{width=0.45\textwidth}
    \begin{tabular}{l|l|c|c}
    \toprule[1pt]
    \multicolumn{1}{c|}{Test Set} & \multicolumn{1}{c|}{Train Set} & AP@0.3 & $\mathrm{T_{all}}$ \\ 
    \midrule[0.5pt]
    \multirow{2}{*}{Sim-R\textsubscript{test}-by-R} & RADDet (R.) & 58.16 & \multirow{2}{*}{0.30s} \\
                      & R. + Sim-R-by-R + Sim-R-by-C & \textbf{68.63} \\ 
    \midrule[0.5pt]
    \multirow{2}{*}{Sim-N\textsubscript{LiDAR}-by-R ($^*$)} & RADDet (R.) & 52.14 & \multirow{2}{*}{0.64s}\\
                      & R. + Sim-R-by-R + Sim-R-by-C & \textbf{60.85} \\ 
    \midrule[0.5pt]
    \multirow{2}{*}{Sim-N\textsubscript{Camera}-by-R ($^*$)} & RADDet (R.) & 48.96 & \multirow{2}{*}{1.29s}  \\
                      & R. + Sim-R-by-R + Sim-R-by-C & \textbf{54.36} \\ 
    \bottomrule[1pt]
    \end{tabular}
    \end{adjustbox}
    \vspace{-2mm}
    \caption{3D Object detection AP@0.3 of RA boxes on simulation data. Sim-R\textsubscript{test}-by-R: generated on RADDet test set scenes with the radar attribute from $\left \{ \sigma, g, Rs, \lambda \right \} _{R}$. Sim-N-by-R: generated on nuScenes v1-mini~\cite{nuscenes} scenes with the radar attribute from $\left \{ \sigma, g, Rs, \lambda \right \} _{R}$. We implement scene perception on nuScenes based on LiDAR and camera separately. $\mathrm{T_{all}}$: runtime per scene for \ourmodel{}. $^*$: more details and results for each category can be found in the Appendix \S~\ref{para: Additional experimental results on unseen sensor and scenarios}.}
    \vspace{-2mm}
    \label{tab:nuscene_ra}
\end{table}

\parahead{3D Object Detection.} We take the 3D radar object detection model RADDet as the baseline. Table~\ref{tab:RC3Dallc_raddet} and Table~\ref{tab:RC3Dallc_carrada} show the results under different training datasets. Unlike in the 2D detection task, 3D detection model trained on real radar data outperforms the model trained entirely on simulation radar data. Nevertheless, models trained using both real and simulation datasets still outperform models trained using only real datasets. These results reveal two important insights: (1) The 3D detection task imposes higher quality requirements on the simulation data because it involves finer 3D features, which are difficult to be fully captured by simulation methods (we suggest that this task should be considered as a challenge for radar simulation). (2) Our simulation data is proved to be valuable from a 3D perspective, which effectively enriches the diversity of radar attributes in the dataset and enhances the performance of existing models on complex tasks. See Fig~\ref{fig:3d_det_sup} in Appendix for more visual results.


Furthermore, for both 2D and 3D detection, we use the simulator trained on Carrada to infer the RADDet scenarios, and the obtained simulation data still improves the model's performance on the RADDet test set, i.e., ``R. + Sim-R-by-C" outperforms ``R.". Swapping the two datasets, the above conclusion still holds. This further demonstrates that the performance gains stem from the diversity of radar attributes rather than overfitting to existing attributes.

\begin{figure}[t!]
    \centering
    \vspace{2mm}
\includegraphics[width=0.49\textwidth]{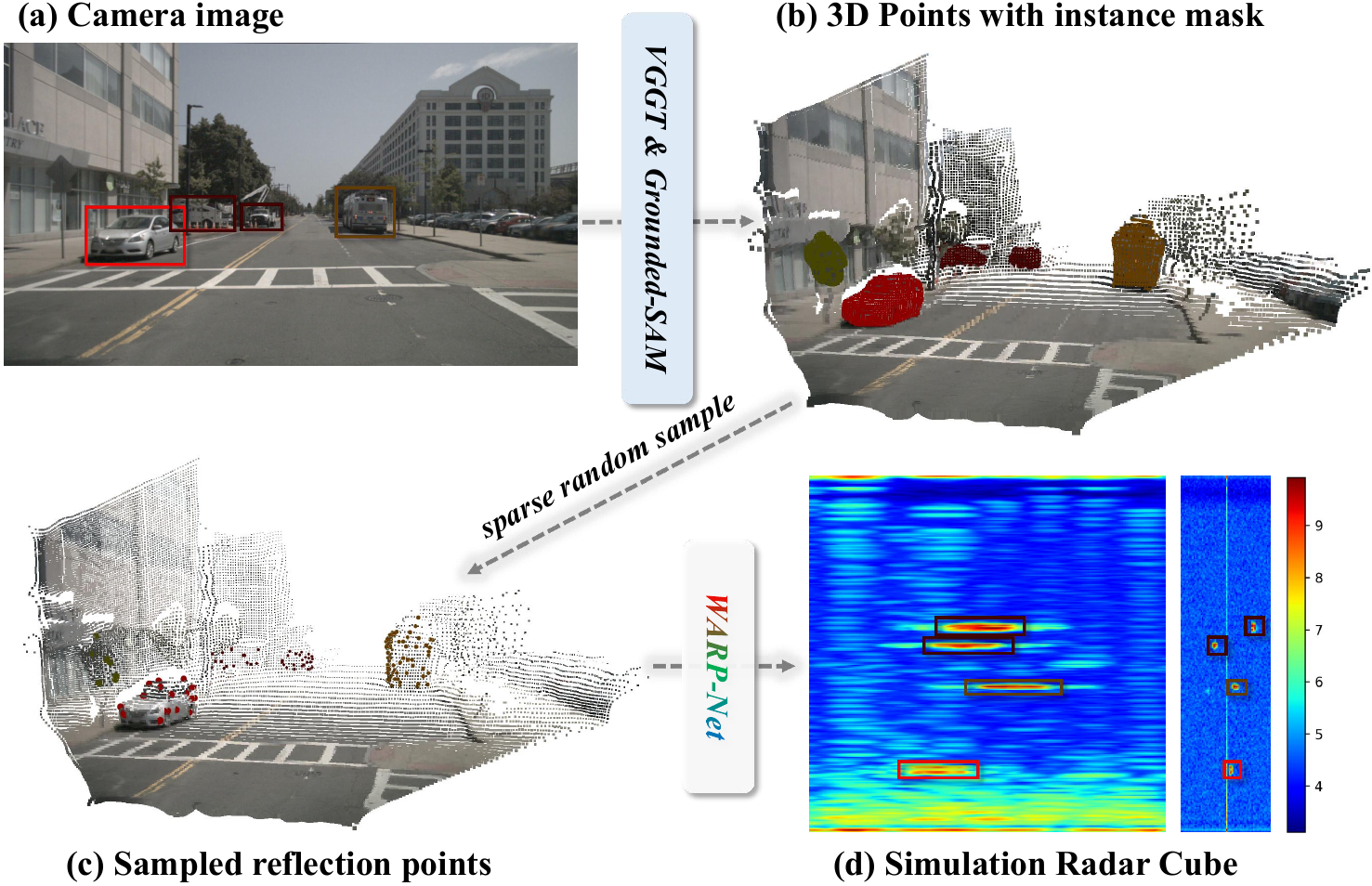}
    \vspace{-6mm}
    \caption{Radar simulation visualization based on nuScenes~\cite{nuscenes} camera data. The colored spheres in subfigure (c) represent the sampled points. See Fig~\ref{fig:sims_camera} in Appendix for more examples.}
    \label{fig:radarcamera}
    \vspace{-6mm}
\end{figure}

\subsection{Simulation based on Other Sensors}

We validate the effectiveness of simulation data under different scene perception methods through downstream models. As shown in Table~\ref{tab:nuscene_ra}, even when employing diverse sensor data for scene perception, models trained solely on real-world data still achieve stable performance on our simulation data. We also evaluate the statistical similarity between these data simulation from other sensors and real radar data. The FID scores for simulation results from Lidar and camera relative to the RADDet test set are 35.8 and 47.1, respectively. These not only demonstrates the physical soundness of our simulation data but also validates the robustness of \ourmodel{} in selecting reflection points. More importantly, this conclusion is verified on an unseen scene (nuScenes~\cite{nuscenes}), providing crucial reference for the feasibility of \ourmodel{} in real-world scenarios. Fig~\ref{fig:radarcamera} visualizes the complete process of how \ourmodel{} achieves radar simulation based on camera data.

Furthermore, in this experiment, the model trained using both real and simulation datasets consistently outperforms models trained only using real data, once again validating the practical value of our simulation data in real-world applications.

\vspace{-1mm}
\section{Conclusion}

\vspace{-2mm}

In this paper, we present \ourmodel{}, a module-complete radar data simulation method designed to simulate radar cubes under definable and controllable radar attributes for arbitrary scenarios. By integrating waveform parameter-based radar attribute embedding, \ourmodel{} can efficiently capture the variations of the radar cube under different radar attributes and accurately simulate radar data that match the real data distribution in range, azimuth and Doppler dimensions. 
Extensive experiments show that our approach significantly enhances the performance of existing models on various downstream tasks and demonstrate  the potential of our approach as a generalized radar data engine for autonomous driving applications.

\parahead{Limitations and Future Work.} The waveform parameter-based representation of radar attributes in \ourmodel{} makes the control of radar attributes simple and intuitive, but it also limits the perception of more detailed radar attributes, which makes it an incomplete substitute for real data in tasks with fine data distributions (e.g., 3D object detection). In addition, the lack of benchmarks makes it difficult to systematically and comprehensively compare our approach with existing methods. Naturally, we hope to address these issues in future work in order to contribute to the advancement of the radar simulation field.

{
    \small
    \bibliographystyle{ieeenat_fullname}
    \bibliography{main}
}

\clearpage
\setcounter{page}{1}
\maketitlesupplementary

\appendix   
\setcounter{table}{0}   
\setcounter{figure}{0}
\setcounter{section}{0}
\setcounter{equation}{0}
\renewcommand{\thetable}{A\arabic{table}}
\renewcommand{\thefigure}{A\arabic{figure}}
\renewcommand{\theequation}{A\arabic{equation}}

In the Supplementary Material, we provide the Appendix (Sec~\ref{appendix}) of this paper along with the project homepage and reproduction code. For additional visualization results and code details, please refer to the \textsc{\textcolor{magenta}{accompanying files}} in the folder.

\section{Appendix}
\label{appendix}
The supplementary material herein extends the discussion and analysis presented in the primary manuscript. It is structured as follows:

\parahead{Introduction to RadSimReal}
(\S~\ref{para: Introduction to Radsimreal})
: This section introduces the RadSimReal~\cite{bialer2024radsimreal} method and discusses its key details.

\parahead{Standard 3D Reflection Signals and Fitting Equations Across Different Datasets}
(\S~\ref{para: Standard 3D reflection Signal on Different Datasets})
: This section shows the geometric consistency of the standard 3D reflection signal across different dimensions in various datasets, and describes how we formulaically model the waveforms of each dimension of the standard 3D reflection signal.

\parahead{Network Structure of WARP-Net}
(\S~\ref{para: Network Structure of WARP-Net})
: This section describes in detail the network architecture of WARP-Net.

\parahead{Datasets}
(\S~\ref{para: Datasets})
: This section describes the RADDet and Carrada datasets that we used in WARP-Net‘s training and the downstream tasks.

\parahead{Formula for Calculating Radar Simulation Quantitative Metrics}
(\S~\ref{para: Formula for Calculating}): This section details the calculation formulas for the metrics in Table~\ref{tab:radartimeandsim}, which are used to assess the similarity between the simulation radar cube and the real radar cube.

\parahead{Mixed Dataset for WARP-Net's Training and Ablation Study}
(\S~\ref{para: Mixed Datasets})
: This section describes the implementation details of the hybrid dataset used to train WARP-Net, and explores the impact of different training data on the simulation performance of WARP-Net through ablation study.

\parahead{More Visualization of the Simulation Data}
(\S~\ref{para: More Visualization of the Simulation Data})
: This section shows more simulation results for the \ourmodel{}.

\parahead{Details of Novel Sensor Trajectory}
(\S~\ref{para: Details of Novel Sensor Trajectory})
: This section describes the implementation details of editing the scene and radar attributes using \ourmodel{}.

\parahead{Training settings and more results for different downstream tasks}
(\S~\ref{para: Training settings and more results for different downstream tasks}):
This section describes the training setup for multiple downstream tasks and additional results.

\parahead{Comparison with RadSimReal and Standard Data Augmentation for Downstream Task}
(\S~\ref{para: Comparison with RadSimReal and Standard Data Augmentation for Downstream Task})
: This section compares the performance of \ourmodel{} with RadSimReal and standard data augmentation for 3D target detection, validating the quality of our simulation data.

\parahead{Additional experimental results on unseen sensor and scenarios}
(\S~\ref{para: Additional experimental results on unseen sensor and scenarios})
: In this section, we proceed to re-evaluate the results presented in Table~\ref{tab:nuscene_ra} of the main paper, focusing on the range-azimuth dimension to further analyze the performance of \ourmodel{} on unseen sensors and scenes. 

\subsection{Introduction to RadSimReal~\cite{bialer2024radsimreal}}
\label{para: Introduction to Radsimreal}
Physical radar simulation~\cite{auer2016raysar, yun2015ray, hirsenkorn2017ray, schussler2021realistic} is a technique for generating a synthetic radar cube ( range-azimuth-Doppler) by physically modeling the environment and radar sensor. The process usually consists of several steps: first, a scene (e.g., a road scene) is generated by a 3D modeling tool (e.g., Unity~\cite{haas2014history} or CARLA~\cite{dosovitskiy2017carla}); then, ray tracing techniques are used to compute the reflections of electromagnetic waves emitted from the radar on objects in the environment; and finally, radar-specific signal processing algorithms are applied to generate the radar cube. The key to this approach is an in-depth understanding of the radar hardware parameters and signal processing algorithms, which makes the simulation process very complex and computationally resource intensive.

Unlike traditional physical radar simulation, RadSimReal realizes radar cube generation via Point Spread Function (PSF).The core idea of RadSimReal is to use PSF to describe the radar sensor's response to reflecting points without the need to deeply understand the radar's specific design details and signal processing process. Specifically, users are able to extract the PSF from a simple set of radar measurements, and integrate the PSF to respond at each reflection point in the scene to quickly generate a high-quality synthetic radar cube (as shown in Equation (1) in the main paper).

For experimental validation, the researcher used RadSimReal to generate synthetic radar cubes containing annotations, and 2D slices of these cubes were used to train a 2D Object Detection model~\cite{zhang2021raddet}. The experimental results show that the detection performance of the 2D Object Detection model trained using the data generated by RadSimReal is comparable to that of the model trained using only real data.

Overall, RadSimReal is inspiring in that modeling radar signals by PSF (generalized in this paper to standard 3D reflection signals) is feasible and can be used for downstream tasks. Based on this, we propose \ourmodel{}, which improves the accuracy and efficiency of radar simulation and realizes the controllability of radar attributes. Meanwhile, we validate the effectiveness of our simulation data on more complex tasks such as 3D Object Detection.

\begin{figure*}[t]
  \centering
  \includegraphics[width=\textwidth]{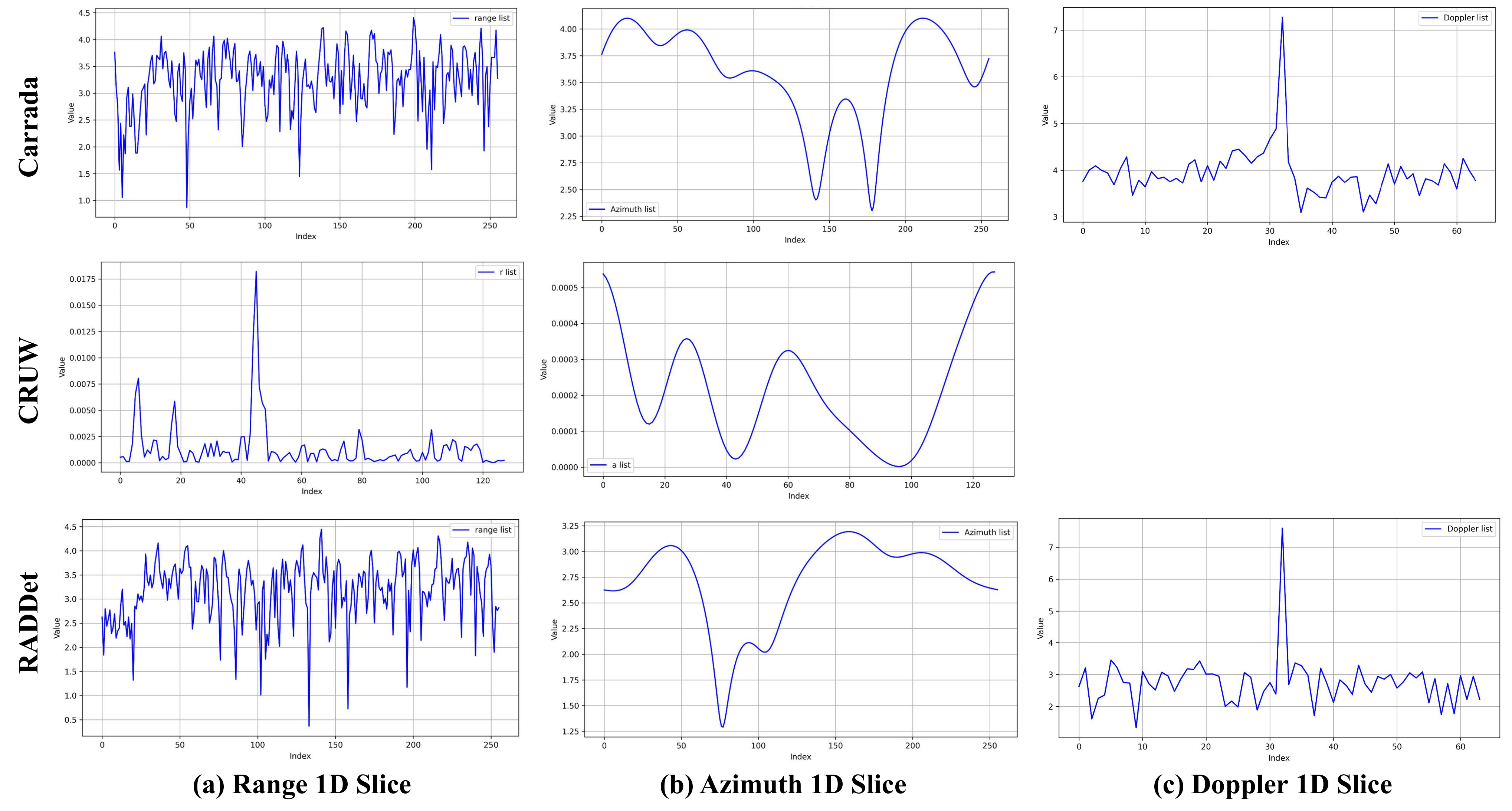}
  \vspace{-4mm}
  \caption{Radar cube 1D slices on different datasets.}
  \vspace{-3mm}
   \label{fig:mdataset1d}
\end{figure*}

\subsection{Standard 3D Reflection Signals and Fitting Equations Across Different Datasets}
\label{para: Standard 3D reflection Signal on Different Datasets}

We show 1D slices of the radar cube on the different datasets in Fig~\ref{fig:mdataset1d}. Since the CRUW dataset~\cite{wang2021rodnet} only provides RA data, we could not get its Doppler slices. Despite the different distributions of the values of the radar cube on the different datasets (because of the different treatments), the standard 3D reflectance signals of each dataset are geometrically very similar in the Range, Azimuth, and Doppler slices, which inspired us to use functions to fit them separately. 

The specific formulas for fitting functions in different dimensions are as follows:
\begin{equation}
\mathbf{S}_{R}(r)=e^{-\frac{(r-r_i)^2}{2\sigma^2 } }
\end{equation}
\begin{equation}
\mathbf{S}_{A}(a)=|FFT((1-p)-p*cos(\frac{2\pi n}{N-1} ))|\otimes \delta (a-a_i)
\end{equation}
\begin{equation}
\mathbf{S}_{D}(d)=g*max\left \{ 1-|d-d_i|,2-4|d-d_i|,0 \right \} 
\end{equation}
where $(r_i,d_i,a_i)$ is the coordinate index of the reflection point in the radar cube, $\sigma$ denotes the standard deviation of the Gaussian function, $FFT$ denotes Fast Fourier Transform, $N$ and $p$ are the length and parameters of the window function, $\otimes$ denotes the convolution operation, and $g$ is the gradient in the segmented linear function. With the above equation, we can further calculate the waveform parameter $\left \{\sigma, g, Rs, \lambda\right \} $ set in this paper to visually characterize the radar attribute.

\subsection{Network Structure of WARP-Net}
\label{para: Network Structure of WARP-Net}
\begin{figure*}[t]
  \centering
  \includegraphics[width=\textwidth]{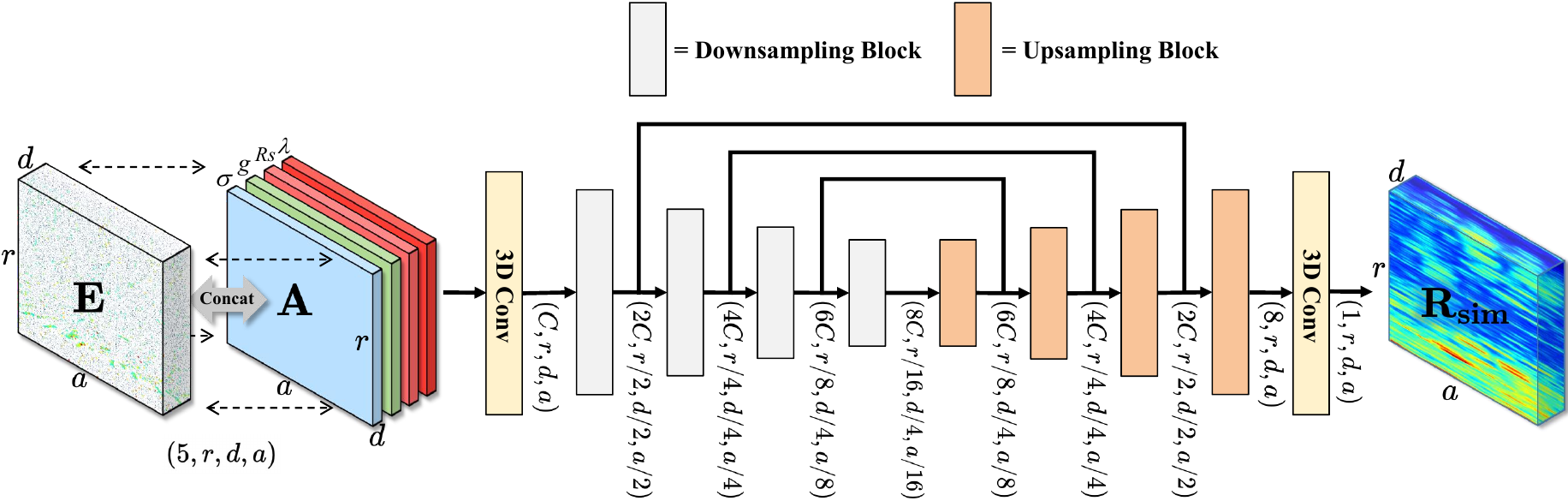}
  \vspace{-4mm}
  \caption{The Architecture of WARP-Net.}
  \vspace{-3mm}
   \label{fig:warp_model}
\end{figure*}

We propose WARP-Net, which directly simulates the radar cube based on the reflection environment tensor $\mathbf{E}$ and the waveform parameter-based radar attribute. Fig~\ref{fig:warp_model} illustrates the architecture of WARP-Net. In order to increase the receptive field in different dimensions (especially the azimuth dimension), we construct the WARP-Net by four downsampling blocks and four upsampling blocks. Each downsampling block consists of two 3D convolutions with output channel numbers of 64, 128, 192, and 256, respectively. The first three upsampling blocks consist of one 3D transpose convolution and three 3D convolutions with output channel numbers of 192, 128, and 64, respectively, while the last upsampling block contains only one 3D transpose convolution and outputs the radar feature map with channel number of 8. In addition, for each 3D convolution and 3D transposed convolution in the WARP-Net, we process its output with a BN layer and a LeakyReLU layer. Finally, we decode the radar feature map using a simple 3 × 3 × 3 3D convolution and the Relu function to obtain the simulated radar cube $\mathbf{R_{\mathrm{sim}}}\in \mathbb{R} ^{ r\times d\times a}$.

\subsection{Datasets}
\label{para: Datasets}
\parahead{RADDet~\cite{zhang2021raddet}.} The RADDet dataset is a publicly available radar dataset intended to be used for multi-class object detection by dynamic road users, and contains radar data denoted as range-azimuth-Doppler tensor and corresponding annotations covering multiple object classes (person, bicycle, car, motorcycle, bus, truck). The dataset was collected via a tat and a pair of stereo cameras, and data capture was performed under clear sky conditions. It contains 10,158 frames of radar data in 3D tensor (256, 256, 64) format containing rich dynamic object information. The training and test sets contain 8,127 and 2,031 frames of radar data, respectively. It should be noted that due to the insufficient number of motorcycle samples, we skip its quantitative assessment, but still consider it across all categories.

\parahead{Carrada~\cite{ouaknine2021carrada}.} Carrada is an automotive radar and camera dataset that provides 12666 frames of data in 30 sequences covering a wide range of scenarios for pedestrians, cars, and bicyclists. The data consists of three forms of annotations: sparse dots, bounding boxes and dense masks, which are suitable for object detection, semantic segmentation and tracking tasks. The training set and test set contain 5,802 and 1,391 frames of usable radar data in 3D tensor (256, 256, 64) format, respectively. Carrada employs a semi-automatic annotation method based on camera information in order to improve efficiency and reduce cost, which also makes its annotation quality poor. We use its training set and validation set together for model training.

\parahead{nuScenes~\cite{nuscenes}.} nuScenes is a large-scale autonomous driving dataset. It comprises 1,000 carefully selected driving scenarios, each lasting 20 seconds, totaling approximately 15 hours of driving data. The dataset covers complex traffic environments in Boston and Singapore. The dataset includes approximately 1.4 million camera images, 390,000 LiDAR scans, 1.4 million radar scans, and 1.4 million object bounding boxes across 40,000 keyframes, suitable for training and testing autonomous driving algorithms. However, since radar scans are provided as sparse point clouds, which filter out weak reflections and noise in the scene, Therefore, we only utilize the Front Camera Images and Lidar Data provided in this dataset for radar simulation. Additionally, due to dataset size constraints, we utilize a subset of this dataset (i.e., v1-mini), containing approximately 400 frames of data.

\subsection{Formula for Calculating Radar Simulation Quantitative Metrics}
\label{para: Formula for Calculating}
In Table~\ref{fig:radarcompare} of the main text, we employed multiple metrics to evaluate the accuracy of the simulated data. Notably, unlike image data, our simulation data $\mathbf{R_{\mathrm{sim}}}\in \mathbb{R} ^{ r\times d\times a}$ constitutes a three-dimensional tensor. Therefore, we apply dimensionality reduction during the computation of FID. This section details their respective calculation formulas. 

\parahead{PPE}, or Per-Pixel Error, directly accounts for the average numerical discrepancy between simulation data and real data, calculated as follows:
\begin{equation}
    \mathrm{PPE} = E(\left | \mathbf{R_{\mathrm{sim}}} - \mathbf{R_{\mathrm{gt}}} \right | )
\end{equation}
where $E(.)$ denotes the average value of all the elements in it, $|.|$ denotes the absolute value calculation, $\mathbf{R_{\mathrm{sim}}}$ is the simulated radar cube, and $\mathbf{R_{\mathrm{gt}}}$ is the ground truth radar cube. 

\parahead{PPSE}, or Per-Point Spectral Error, evaluates the average error between simulation data and real data from the frequency domain perspective. Its calculation formula is as follows:
\begin{equation}
    \mathrm{PPSE} = E(\left | FFT(\mathbf{R_{\mathrm{sim}}}) - FFT(\mathbf{R_{\mathrm{gt}}}) \right | )
\end{equation}
where $FFT$ denotes the Fourier Transform.

\parahead{FID} is a common metric used to assess statistical similarity between images. Therefore, we first reduce the dimensionality of the radar cube by taking the maximum Doppler dimension to obtain the range-azimuth image, as follows:
\begin{equation}
    \mathbf{I_{\mathrm{sim}}} = \max(\mathbf{R_{\mathrm{sim}}}, \text{ axis}=1)
\end{equation}
\begin{equation}
    \mathbf{I_{\mathrm{gt}}} = \max(\mathbf{R_{\mathrm{gt}}}, \text{ axis}=1)
\end{equation}
where $\mathbf{I_{\mathrm{sim}}}$ and $\mathbf{I_{\mathrm{gt}}}$ are radar images obtained by dimensionality reduction from $\mathbf{R_{\mathrm{sim}}}$ and $\mathbf{R_{\mathrm{gt}}}$, respectively. We then compute their FID metrics as follows:
\begin{equation}
\text{FID} = \|\mu_{\mathrm{sim}} - \mu_{\mathrm{gt}}\|^2 + \text{Tr}(\Sigma_{\mathrm{sim}} + \Sigma_{\mathrm{gt}} - 2 \sqrt{\Sigma_{\mathrm{sim}} \Sigma_{\mathrm{gt}}})
\end{equation}
where $\mu_{(.)}$ denotes the mean vector of features for the radar images, $\Sigma_{(.)}$ denotes the feature covariance matrix for the radar images, and $\text{Tr}(.)$ denotes the trace of the matrix.

\subsection{Mixed Dataset for WARP-Net's Training and Ablation Study}
\label{para: Mixed Datasets}

\begin{figure*}[t]
  \centering
  \includegraphics[width=\textwidth]{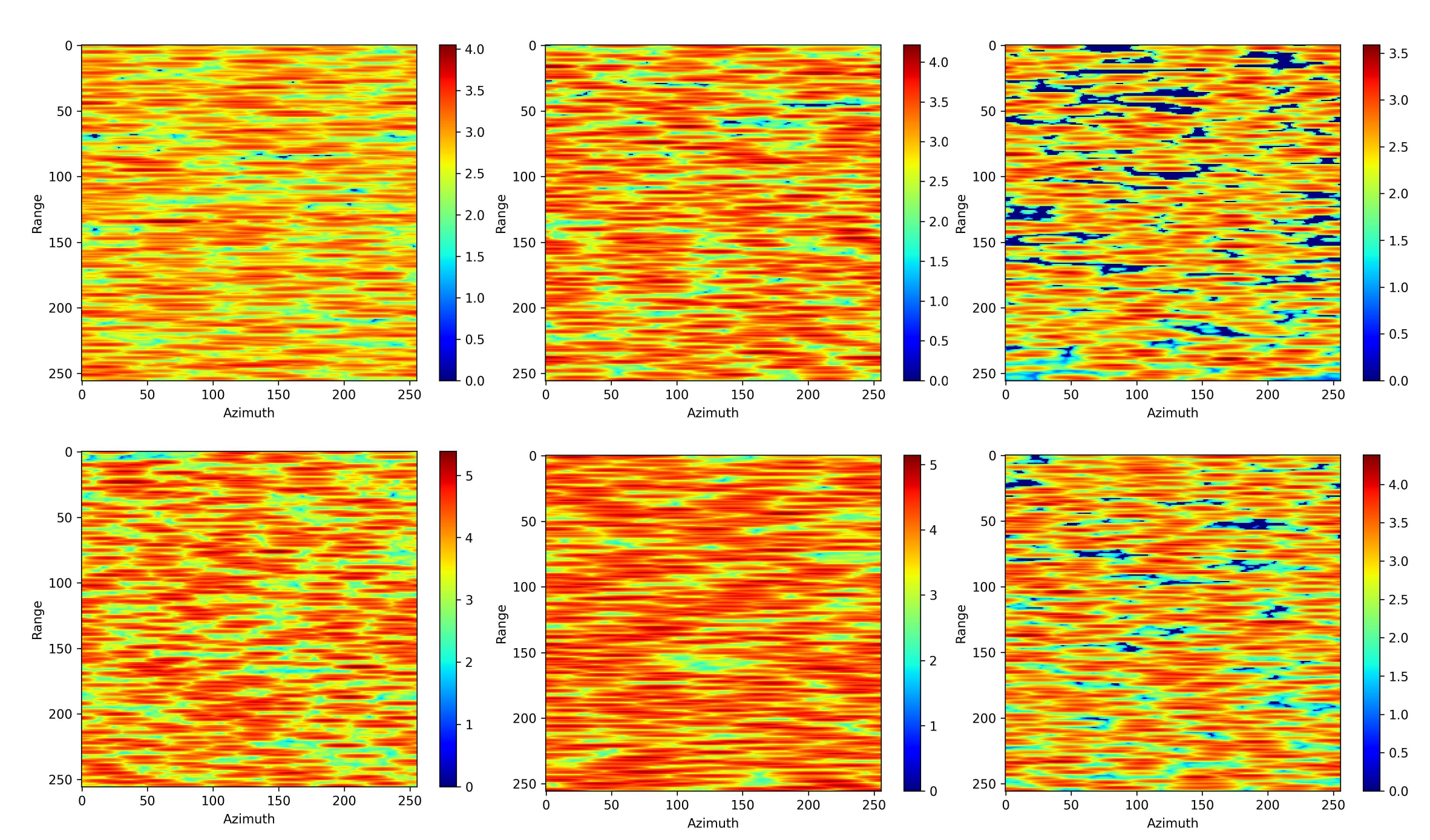}
  \caption{Simulation results of different standard 3D reflection signals at random noise reflection points.}
   \label{fig:radarsimnoise}
\end{figure*}

\parahead{Implementation Details.}
In this paper, we produce a mixed dataset rich in radar attributes to train WARP-Net. As we mentioned in the main paper, the mixed dataset is divided into three parts: the real dataset A, the synthetic dataset B, and the simulation dataset C. The details of the practice are described below.

Since there are fewer real radar cube datasets in the form of range-azimuth-Doppler tensor, we only use the training sets of RADDet~\cite{zhang2021raddet} and Carrada~\cite{ouaknine2021carrada} in practice to compose the real dataset A. Therefore, the real dataset A covers only two radar attributes.

For the production of the synthetic dataset B, we combine the method in RadSimReal~\cite{bialer2024radsimreal} with the three fitting functions demonstrated in Sec~\ref{para: Standard 3D reflection Signal on Different Datasets}. We use the method in RadSimReal to produce the synthetic dataset B. The synthetic dataset B is a set of the three fitting functions. Specifically, we construct the set of fitting functions by randomly selecting the fitting parameters from the following set: $\sigma\in \left \{ 2.4,2.5,2.6,2.7,2.8 \right \} $, $N \in \left \{ 6,7,8,9,10 \right \} $, $g \in \left \{ 0.5, 0.6, 0.7\right \} $ and $p \in \left \{ 0.1, 0.2, 0.3 \right \} $. Note that we measure the average of the fitting parameters in the RADDet training set, which are: $\sigma\in=2.6$, $N=8$, $g=0.6$ and $p=0.1$. We then use these fitting functions to compute the standard 3D reflection signals according to Eq. (2) in the main paper (Fig~\ref{fig:radarsimnoise} illustrates these signals on the random noise reflection points). Finally, we generate the radar data under these standard 3D reflection signals (randomly selected standard 3D reflection signals) in the scenario of the RADDet training set according to Eq. (1) in the main paper. To summarize, the number of samples in the synthetic dataset B is equal to the number of scenes in the RADDet training set, but the radar attributes are much richer.

For simulation dataset C, due to the lack of real datasets, we only train two 3D U-Nets with radar attribute embeddings removed on RADDet and Carrada. We use the two 3D U-Nets (with radar attribute embeddings removed) trained on RADDet and Carrada to inference on each other's training set scenarios separately, and we finally obtain a simulated dataset C, which has the same number of samples as the real dataset.

Finally, since the ratio of train sets between RADDet and Carrada is approximately 4: 3, the overall ratio of datasets A, B and C is roughly 7: 4: 7.

\parahead{Ablation Study.}
We explore the effect of training data type on the performance of WARP-Net. As we do in the main paper, we extract all reflection points and reflection intensities (both scene and noise) from a real radar cube using the CFAR algorithm. Then, we simulate the radar cube using WARP-Net. Table~\ref{tab:ablationABC} records the simulation errors of WARP-Net with different training data. Considering only the simulation quality, WARP-Net trained only with the real dataset A is slightly better than WARP-Net trained only with the simulation dataset C, and much better than WARP-Net trained only with the synthetic dataset B.

\begin{table}[t]
    \centering
    \setlength{\tabcolsep}{3pt}
    \begin{adjustbox}{width=0.46\textwidth}
    \begin{tabular}{ccc|cc|cc}
        \toprule[1pt]
        \multirow{2}{*}{Dataset A} & \multirow{2}{*}{Dataset B} & \multirow{2}{*}{Dataset C}  & \multicolumn{2}{c|}{RADDet~\cite{zhang2021raddet}} & \multicolumn{2}{c}{Carrada~\cite{ouaknine2021carrada}} \\
         & & & $\mathrm{PPE}$ & $\mathrm{PPE}_{s}$ & $\mathrm{PPE}$ & $\mathrm{PPE}_{s}$ \\
        \midrule[0.5pt]
        \cmark & & &  \underline{0.271} & \textbf{0.008} & \textbf{0.270} & \textbf{0.009} \\
        & \cmark & & 0.480 & 0.138 & 0.505 & 0.171 \\
        & & \cmark & 0.363 & 0.033 & 0.487 & 0.027 \\
        \midrule[0.5pt]
        \cmark & \cmark & \cmark & \textbf{0.267} &  \underline{0.009} &  \underline{0.271} &  \underline{0.012} \\
        \bottomrule[1pt]
    \end{tabular}
    \end{adjustbox}
    \caption{Comparison of WARP-Net simulation quality under different types of training data. \textbf{Bold}: Best. \underline{Underline}: Second.}
    \label{tab:ablationABC}
\end{table}

These results reveal two important insights: 1. The real dataset A significantly improves the simulation quality of WARP-Net. 2. Although the synthetic dataset B effectively enriches the radar attributes of the mixed dataset, it damages the simulation quality of WARP-Net (thus we use the simulation dataset C to mitigate this negative effect and further enrich the diversity of the mixed dataset).

\subsection{More Visualization of the Simulation Data}
\label{para: More Visualization of the Simulation Data}

\begin{figure*}[t]
  \centering
  \includegraphics[width=\textwidth]{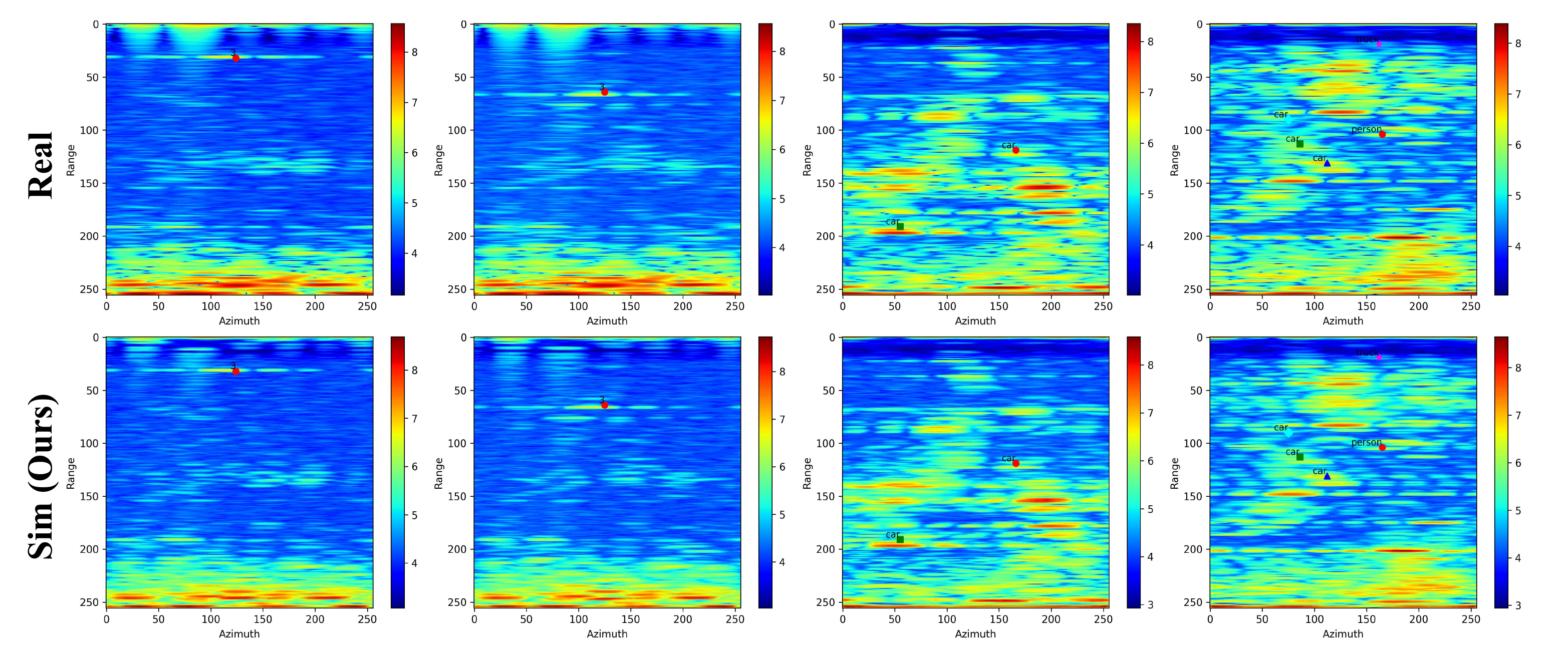}
  \caption{Simulation results on the RADDet and Carrada test sets.}
   \label{fig:moresim}
\end{figure*}

Fig~\ref{fig:moresim} presents additional simulation results of \ourmodel{} under the same conditions as Fig~\ref{fig:radarcompare} in the main text. It can be observed that our simulation data exhibits high similarity to real-world data.

\begin{figure*}[t]
  \centering
  \includegraphics[width=\textwidth]{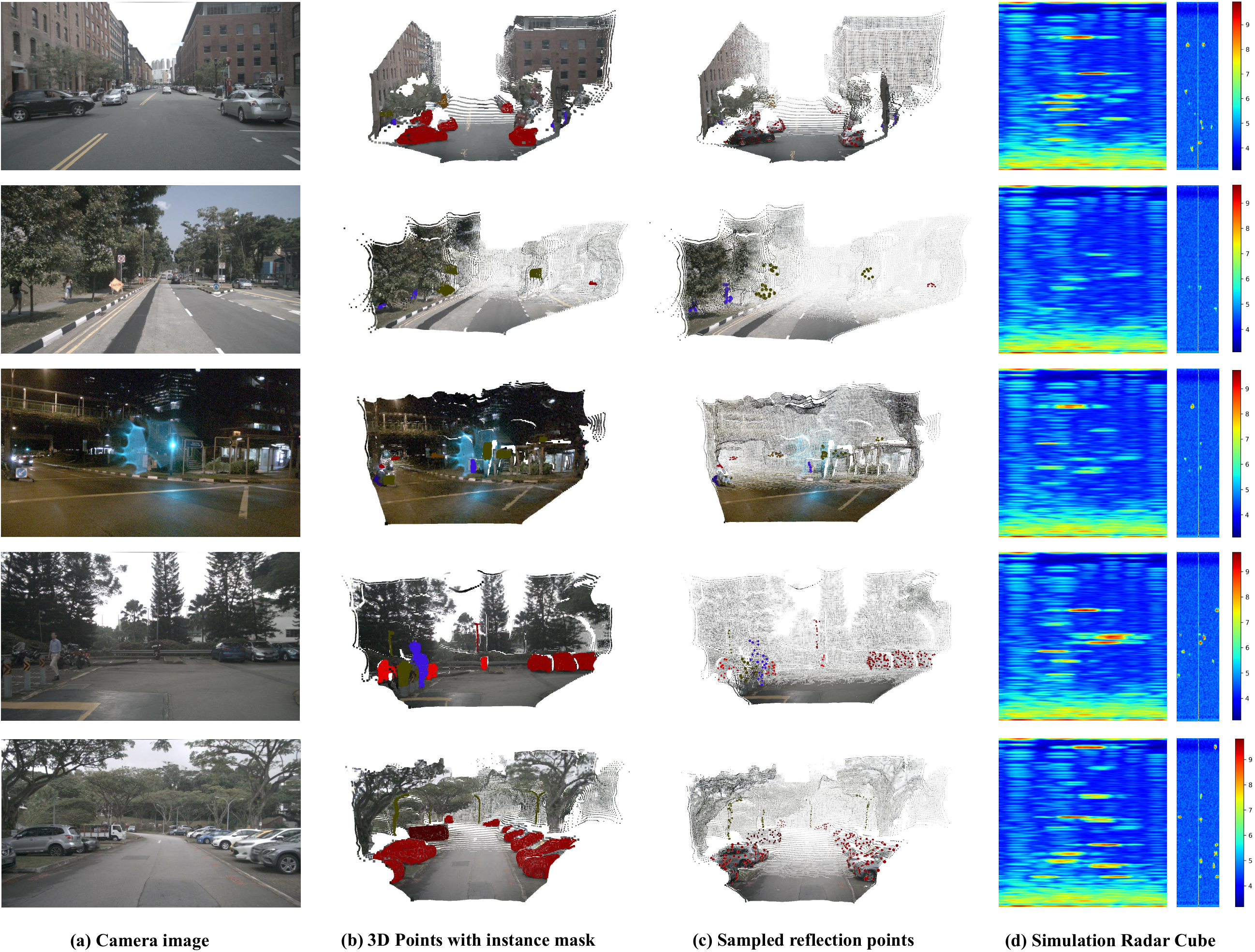}
  \caption{Radar simulation results based on camera images. Note that since the dataset does not provide precise speed information, we randomly assigned speeds to different instances in the simulation, which does not affect the validation on downstream tasks.}
   \label{fig:sims_camera}
\end{figure*}

Furthermore, we present additional radar simulation results based on camera images in Fig~\ref{fig:sims_camera} as a supplement to Fig~\ref{fig:radarcamera} in the main text. Objectively speaking, \ourmodel{}'s results are inevitably influenced by scene reconstruction quality. For instance, inaccuracies in target positions within the 3D point cloud may introduce errors in simulation results. Nevertheless, our simulations consistently yield visually plausible outcomes and remain directly usable for downstream tasks, fully demonstrating \ourmodel{}'s robustness.

\subsection{Details of Novel Sensor Trajectory}
\label{para: Details of Novel Sensor Trajectory}

\begin{figure*}[t]
  \centering
  \includegraphics[width=\textwidth]{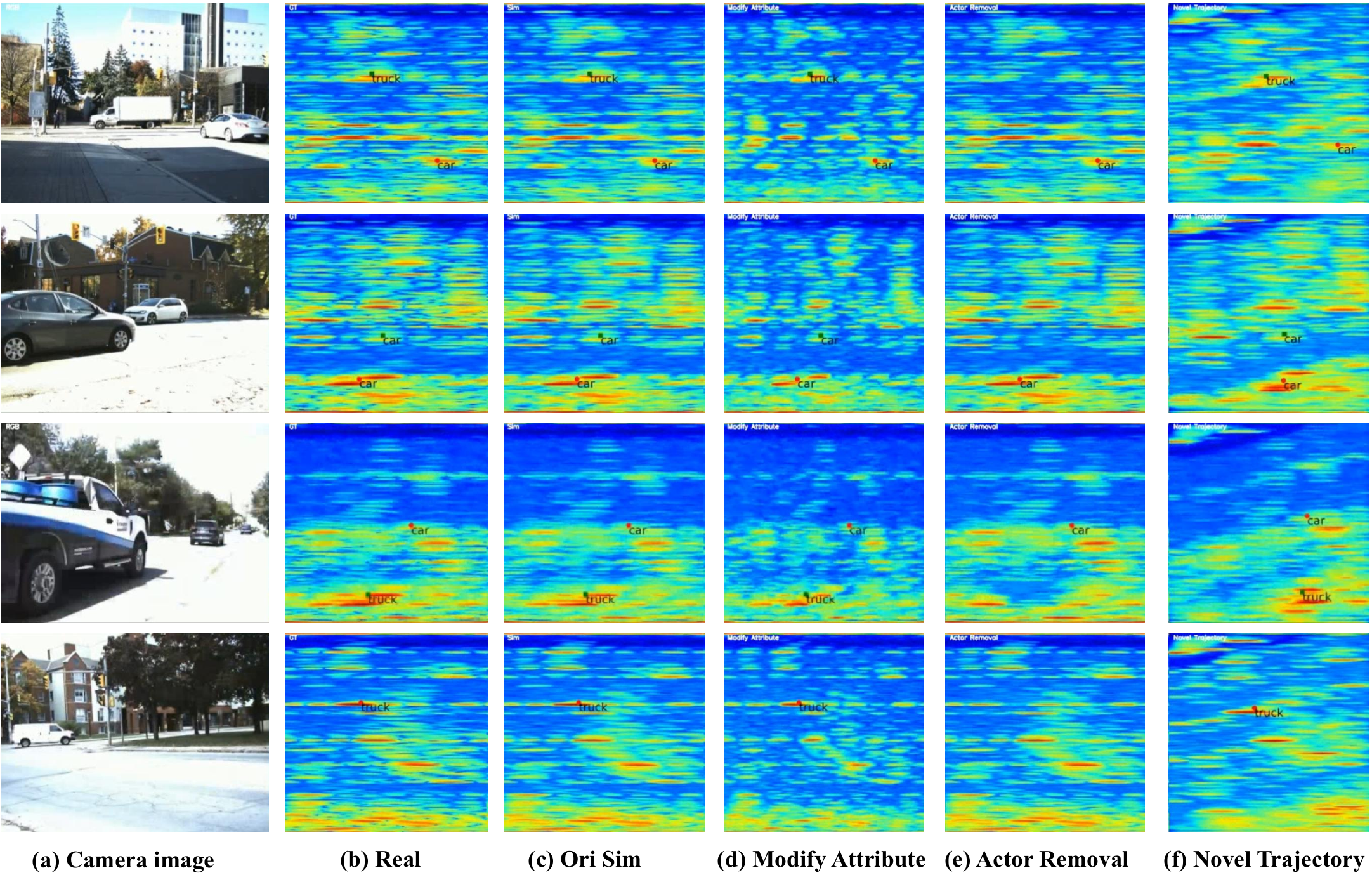}
  \vspace{-6mm}
  \caption{More results of radar simulation and scene editing.}
   \label{fig:morenvs}
\end{figure*}

Our \ourmodel{} allows editing of the scene and editing of the simulation radar attributes by modifying the reflection environment tensor $\mathbf{E}$ and radar attribute embedding. The visualization results are shown in Fig~\ref{fig:morenvs}. The following are the specific implementation details of each editing operation.

\parahead{Modify Attribute.} No need for complicated operation, we can directly modify the waveform parameters $\left \{\sigma, g, Rs, \lambda\right \}$ which are input to WARP-Net to simulate different radar cube with different radar attributes for the same scene.

\parahead{Novel Trajectory.} As described in Sec~\ref{sec:es}, the reflection environment tensor $\mathbf{E}$ in \ourmodel{} corresponds dimensionally to the simulated radar cube. In other words, each of the three dimensions of the reflection environment tensor $\mathbf{E}$ has a specific physical meaning, i.e., range, azimuth, and Doppler. Therefore, we are free to edit the viewpoint in the scene perception. Then, the new reflection environment tensor $E_{\mathrm{new}}$ is obtained by recalculating the corresponding coordinates of the scene reflection points in the reflection environment tensor under the new viewpoint. Finally, we use WARP-Net to process the reflective environment tensor $\mathbf{E_{\mathrm{new}}}$ to obtain the simulated radar cube under the new viewpoint.

\parahead{Actor Removal.} We realize the removal function by directly modifying the reflection intensity of the reflection points on the actor. Specifically, for the actor to be removed in the scene, we can box its reflection points by 3D Object detection, and then reduce the reflection intensity of its reflection points to be similar to that of the noise reflection points. Then, we directly use WARP-Net to process the intensity-modified reflection environment tensor to get the simulated radar cube of actor removal.

\subsection{Training settings and more results for different downstream tasks}
\label{para: Training settings and more results for different downstream tasks}

\begin{table*}[t]
    \centering
    \setlength{\tabcolsep}{4pt}
    \setlength{\abovecaptionskip}{-2pt}
    \setlength{\belowcaptionskip}{2pt}
    \renewcommand{\arraystretch}{1.1}
    \footnotesize
    \begin{adjustbox}{max width=\linewidth}
    \begin{tabular}{l|ccccc|ccccc}
        \toprule[1pt]
         \multicolumn{1}{c|}{\multirow{2.5}{*}{Train Set}} & \multicolumn{5}{c|}{IoU (\%) of RA / RD} & \multicolumn{5}{c}{Dice (\%) of RA / RD}  \\
         \cmidrule(lr){2-11} & Bkg. & Ped. & Cyclist & Car & All & Bkg. & Ped. & Cyclist & Car & All  \\
        \midrule[0.5pt]
         Carrada (C.) & 99.7 / 99.7 & 12.1 / 23.7 & \textbf{15.7} / 38.1 & 29.2 / 56.3 & 39.2 / 54.4 & 99.8 / 99.8 & 21.6 / 38.3 & \textbf{27.1} / 55.1 & 45.2 / 72.0 & 48.5 / 66.3 \\
         Sim-C-by-C + Sim-C-by-R & 99.8 / 99.7 & 8.4 / 46.3 & 10.3 / 38.9 & 29.3 / 46.7 & 37.0 / 57.9 & 99.9 / 99.8 & 15.5 / 63.3 & 18.8 / 56.0 & 45.3 / 63.7 & 44.9 / 70.7 \\
         C. + Sim-C-by-C & 99.8 / 99.8 & \textbf{18.7} / 52.4 & 14.5 / 40.6 & 31.8 / \textbf{67.5} & \textbf{41.2} / 65.1 & 99.9 / 99.9 & \textbf{31.5} / 68.8 & 25.3 / 57.7 & 48.3 / \textbf{80.6} & \textbf{51.2} / 76.7 \\
         C. + Sim-C-by-R & 99.8 / 99.8 & 14.7 / 54.1 & 14.8 / 33.8 & 29.2 / 66.0 & 39.6 / 63.4 & 99.9 / 99.9 & 25.6 / 70.2 & 25.8 / 50.5 & 45.2 / 79.5 & 49.1 / 75.0 \\
         C. + Sim-C-by-C + Sim-C-by-R & 99.8 / 99.8 & 14.5 / \textbf{54.9} & 14.6 / \textbf{44.4} & \textbf{33.4} / 64.7 & 40.2 / \textbf{66.2} & 99.9 / 99.9 & 25.7 / \textbf{71.3} & 25.4 / \textbf{61.6} & \textbf{51.4} / 76.1 & 49.7 / \textbf{77.9} \\
        \bottomrule[1pt]
    \end{tabular}
    \end{adjustbox}
    \vspace{3mm}
    \caption{Performance comparison of multi-view semantic segmentation on the Carrada test set. The baseline is MVRSS~\cite{ouaknine2021multi}, which analyzes RA and RD views of continuously acquired radar signals to semantically segment them.}
    \label{tab:RCseg_all}
\end{table*}

\parahead{Multi-View Radar Semantic Segmentation.}
In this paper, we use MVRSS to verify the semantic accuracy and stability of \ourmodel{} simulation data. We present the complete results on the Carrada test set in Table~\ref{tab:RCseg_all}, as a supplement to Table~\ref{tab:RCseg} in the main text. For model training, we only modified the number of training epochs in the default training configuration: from training 300 epochs to training 200 epochs, thus reducing the time cost. Nevertheless, the model trained jointly on our simulation data and real data still outperforms the model trained with 300 epochs in the original paper.

\begin{table}[t]
    \centering
    \resizebox{\columnwidth}{!}{
    \begin{tabular}{c|c|c|c}
        \toprule[1pt]
        Train Set & Amounts & Train Set & Amounts   \\
        \midrule[0.5pt]
        RADDet (R.) & 22694 & Carrada (C.) & 5802 \\
        Sim-R-by-R + Sim-R-by-C & 45388 & Sim-C-by-C + Sim-C-by-R & 11604 \\
        R. + Sim-R-by-R & 45388 & C. + Sim-C-by-C & 11604 \\
        R. + Sim-R-by-C & 45388 & C. + Sim-C-by-R & 11604 \\
        R. + Sim-R-by-R + Sim-R-by-C & 68082 & C. + Sim-C-by-C + Sim-C-by-R & 17406 \\
        \midrule[0.5pt]
        \midrule[0.5pt]
        Test Set & Amount & Test Set & Amount   \\
        \midrule[0.5pt]
        RADDet (R.) & 5619 & Carrada (C.) & 1399 \\
        \bottomrule[1pt]
    \end{tabular}
    }
    \caption{The amount of RD slices in the training and test sets for the RADDet and Carrada datasets. For the RADDet dataset, we extract Range-Doppler (RD) slices by selecting the maximum value along the angle dimension. For the Carrada dataset, we determine the RD slices based on the bounding box dimensions of the "car" class, selecting the maximum value along the angle dimension accordingly.}
    \label{tab:2d_data}
    \vspace{-4mm}
\end{table}

\begin{figure*}[t]
  \centering
  \includegraphics[width=1\textwidth]{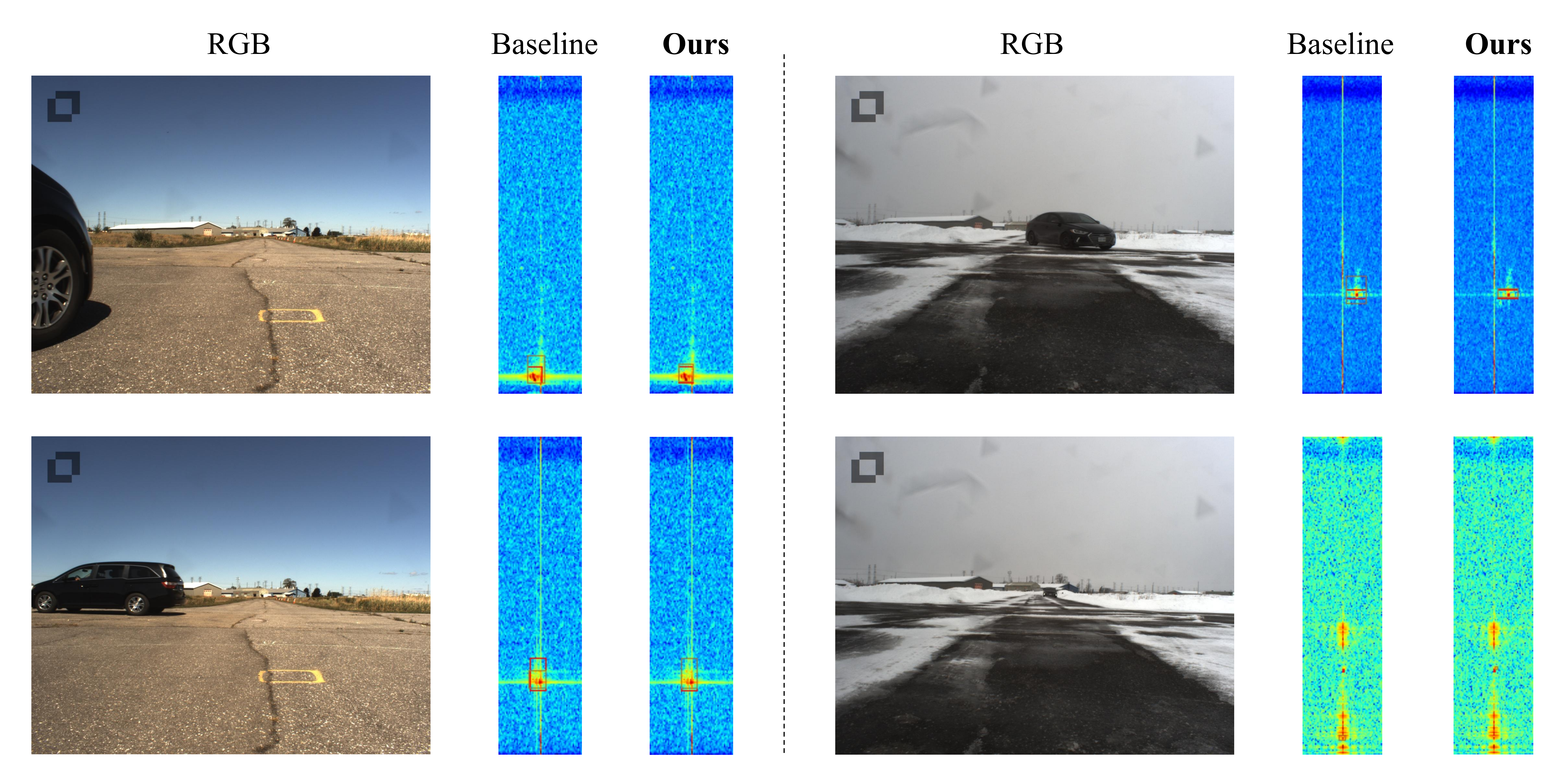}
  \caption{Visualization of 2D slices (range-Doppler) of radar cube from Carrada. The areas on both sides of the dashed line present RGB images from different scenes, along with the detection results on RD slices from both the baseline and our model (Ours). Specifically, the baseline model is trained only on real data, while our model is co-trained on our simulation data and real data. In each of the RD slice, red bounding boxes denote the ground truth, while brown bounding boxes represent the predictions.}
   \label{fig:2d_carrada}
\end{figure*}

\parahead{2D Object Detection.} We conduct experiments on the RADDet and Carrada datasets for 2D object detection using Range-Doppler (RD) slices. For our object detection model, we adopt RTMDet-Tiny \cite{lyu2022rtmdet}, which is trained on a single NVIDIA A800 GPU. Specifically, The model is trained for 100 epochs on the RADDet dataset and 300 epochs on the Carrada dataset, using the AdamW optimizer with a cosine annealing learning rate schedule. We initialize our model with COCO pre-trained weights and evaluate performance on annotated RD slices. Table~\ref{tab:2d_data} presents the number of RD slices in the training and test sets for each dataset. Finally, we select the best-performing models and evaluate their performance using bbox AP@0.5:0.95 (AP), AP@0.5, and AP@0.75.
 
For dataset selection, we adopt different strategies based on dataset characteristics. In the RADDet dataset, following 3D detection approaches, we select five representative object categories, including car, bicycle, person, bus and truck. In contrast, for Carrada, due to its poor annotation quality, we regenerate new annotations \cite{zhang2023peakconv} for each scene and focus on the representative car category for detection. 

Fig~\ref{fig:2d_carrada} provides a comparison of some of the 2D detection results on the Carrada. Our simulation data from \ourmodel{} can effectively enhance the detection performance of the model.

\parahead{3D Object Detection.}
In this paper, we use the RADDet model as a baseline in the 3D detection task. We follow its default training settings. Specifically, we train the model with a batch size of 3 for 1000 epochs using the Adam optimizer. The initial learning rate is 0.0001; after 60K warm-up steps, the learning rate decays to 0.96 every 10K steps. The NMS thresholds for the 3D and 2D detection headers are 0.1 and 0.3, respectively. In fact, we find that the model has largely converged or even started to overfit after being trained for about 300 epochs, so we save the best-performing weights for comparison. 

\begin{table*}[t]
    \small
    \centering
    \vspace{2mm}
    \resizebox{2\columnwidth}{!}{
    \begin{tabular}{l|l|c|c|c|c|c|c}
        \toprule[1pt]
        \multirow{2}{*}{Test Set} & \multicolumn{1}{c|}{\multirow{2}{*}{Train Set}} & \multicolumn{6}{c}{AP@0.3  of 
 RAD / RA}  \\
         \cline{3-8} & & \multicolumn{1}{c|}{Pedestrian} & \multicolumn{1}{c|}{Cyclist} & \multicolumn{1}{c|}{Car} & \multicolumn{3}{c}{All} \\
        \midrule[0.5pt]
        \multirow{5}{*}{Carrada} & Carrada (C.) & 4.49 / \underline{25.62}  & \underline{4.40} / 6.29 & 21.19 / 42.84 & \multicolumn{3}{c}{9.83 / 29.91} \\
        & Sim-C-by-C + Sim-C-by-R & 3.66 / 23.78 & 3.14 / \textbf{13.84} & 15.95 / 42.55 & \multicolumn{3}{c}{7.68 / 29.19 } \\
        & C. + Sim-C-by-C & 8.23 / 25.15 & 2.52 / 13.21 & 20.90 / \textbf{46.67} & \multicolumn{3}{c}{12.10 / 31.39 } \\
        & C. + Sim-C-by-R & \underline{8.78} / 23.50 & \textbf{8.49} / 10.06 & \textbf{21.84} / 40.47 & \multicolumn{3}{c}{13.05 / 28.25 } \\
        & C. + Sim-C-by-C + Sim-C-by-R & \textbf{10.84} / \textbf{27.44} & \underline{4.40} / 6.92 & 19.45 / 44.59 & \multicolumn{3}{c}{$\textbf{13.21}_{\textcolor{red}{\textbf{+3.38}}}$ / {$\textbf{31.55}_{\textcolor{red}{\textbf{+1.64}}}$}} \\
        \bottomrule[1pt]
    \end{tabular}
    }
    \vspace{-2mm}
    \caption{3D Object detection AP@0.3 on Carrada. The baseline is the RADDet model~\cite{zhang2021raddet}, which takes as input a 3D radar cube and predicts 2D boxes via a RA YOLO Head and 3D boxes via a RAD YOLO Head.}
    \label{tab:RC3Dallc_all}
\end{table*}

\begin{figure*}[t]
  \centering
  \includegraphics[width=1.0\textwidth]{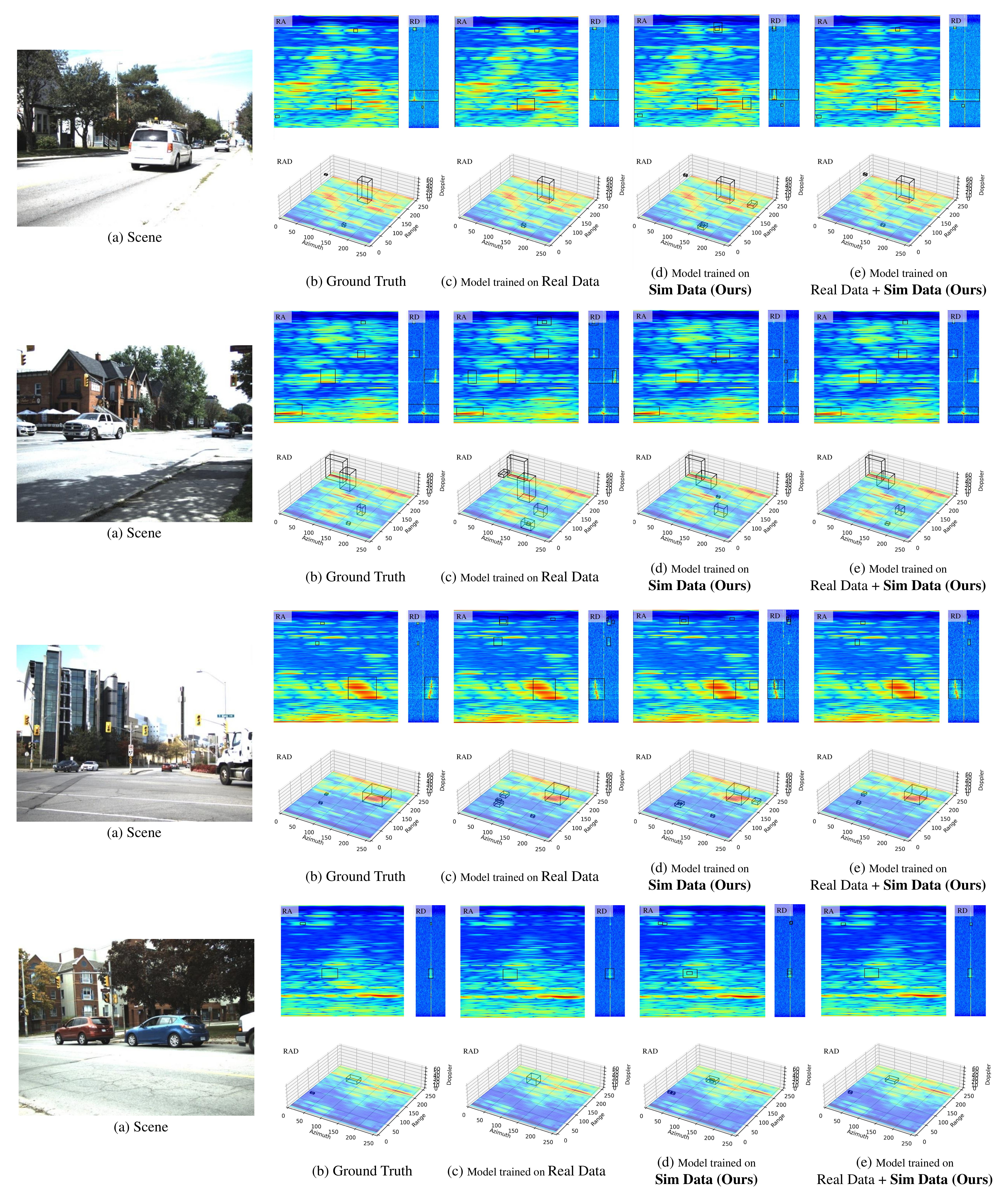}
  \caption{Supplementary 3D detection results in the RADDet test set.}
   \label{fig:3d_det_sup}
\end{figure*}

Similarly, we present the complete results on the Carrada test set in Table~\ref{tab:RC3Dallc_all}, as a supplement to Table~\ref{tab:RC3Dallc_carrada} in the main text. Nevertheless, the visualization results of the experiments are shown in Fig~\ref{fig:3d_det_sup}.

\subsection{Comparison with RadSimReal and Standard Data Augmentation for Downstream Task}
\label{para: Comparison with RadSimReal and Standard Data Augmentation for Downstream Task}

To further validate the quality of our simulation data, we compare the performance of \ourmodel{} with RadSimReal and standard data augmentation on 3D object detection. Table~\ref{tab:CSRS} shows the comparison results on RADDet. Among the different radar simulation methods, although the model trained on the simulation data of RadSimReal gains a slight performance gain, it is still not as good as the model trained on the simulation data of \ourmodel{}. This fully illustrates the high qualitative requirements of training data for 3D object detection, and effectively proves the physical rationality and accuracy of our \ourmodel{}'s simulation data. 

Furthermore, the application of Gaussian noise and random rotation in the training leads to a performance loss for the detection model, as both destroy the physically relevant geometric representations in the radar data. We point out that, while the radar cube visually exhibits some randomness in the range and Doppler dimensions (which it is not), it is strictly smooth and morphologically identical in the azimuth dimension. Therefore, the diversity of the radar dataset cannot be enriched by simple data augmentation, which is quite different from the RGB image dataset.

\begin{table*}[]
    \centering
    \small
    \setlength{\tabcolsep}{2.8pt}
    \begin{tabular}{@{}cccccccc@{}}
    \toprule
    \multirow{2}{*}{Test set}                                & \multirow{2}{*}{Train set}                & \multicolumn{6}{c}{AP@0.3}                       \\ \cmidrule(l){3-8} 
                                                             &                                           & person & bicycle & car   & bus   & truck & all   \\ \midrule
    \multicolumn{1}{c|}{\multirow{5}[7]{*}{RADDet 3D test set}} & \multicolumn{1}{c|}{RADDet (R.)}           & 32.91 & 22.25 & 65.42 & 43.42 & 51.55  & 55.50  \\ \cmidrule(l){2-8} 
    \multicolumn{1}{c|}{}                                    & \multicolumn{1}{c|}{R. + RadSimReal}      & 33.76  & 22.25   & 65.20 & 38.16 & 54.99 & 55.83 \\ \cmidrule(l){2-8} 
    \multicolumn{1}{c|}{}                                    & \multicolumn{1}{c|}{R. + Gaussion Noise}  & 31.97  & 17.34   & 62.31 & 44.74 & 51.81 & 53.05 \\ \cmidrule(l){2-8} 
    \multicolumn{1}{c|}{}                                    & \multicolumn{1}{c|}{R. + Random Rotation} & 30.42  & 16.81   & 65.62 & 28.19 & 47.55 & 54.13 \\ \cmidrule(l){2-8} 
    \multicolumn{1}{c|}{}                                    & R.+ Sim-on-R-by-R + Sim-on-R-by-C         & \textbf{36.85}  & \textbf{27.75} & \textbf{69.25} & \textbf{44.74} & \textbf{57.64} & \textbf{59.66} \\ \bottomrule
    \end{tabular}
    \caption{Comparison of \ourmodel{} with RadSimReal and Standard Data Augmentation on 3D Object Detection. \textbf{Bold}: Best}
    \label{tab:CSRS}
\end{table*}

\subsection{Additional experimental results on unseen sensor and scenarios}
\label{para: Additional experimental results on unseen sensor and scenarios}

We present the complete quantitative results of the 3D detection model on simulated data in Table~\ref{tab:nuscene}. Additionally, we display the corresponding results from the RA perspective in Table~\ref{tab:nuscene2d}. Specifically, we project the 3D bounding boxes of the detection model onto the range-azimuth dimension and then perform NMS processing to obtain 2D bounding boxes from the RA perspective.

Intuitively, significant performance variations exist across different datasets for 3D detection results of various categories. However, after grouping categories like car, bus, and truck into a single category, the quantitative results show a marked improvement. This indicates that errors in simulation data primarily stem from ambiguous object categorization. This is because the distinction between different object categories in radar simulations mainly lies in the distribution of reflection points, rather than reflection intensity (e.g., cars and buses exhibit similar reflection intensities).

In contrast, from the RA perspective, the differences in results across categories and datasets diminish. This is because the distribution of reflection points in Doppler is directly related to the target's radial velocity relative to the sensor, involving both relative position and relative velocity. However, lacking this information, we can only assign approximate Doppler values to the point cloud of each target. Consequently, when this dimension is ignored (considering only range and azimuth), quantitative metrics improve significantly. Similarly, treating cars, buses, and trucks as a single category further reduces differences across datasets.

Overall, we arrive at a crucial conclusion: even when starting from camera images or lidar point clouds, our method robustly simulates physically plausible radar data, though it remains influenced by the selection of reflected points—a key challenge faced by previous radar point cloud simulation approaches. However, from the perspective of practical autonomous driving applications, this influence appears less critical, as we need only determine whether obstacles exist in the scene, not their specific category or velocity.

\begin{table*}[t]
    \centering
    \vspace{2mm}
    \begin{adjustbox}{width=\textwidth}
    \begin{tabular}{c|c|c|cccc|c}
    \toprule[1pt]
    Test Set & Train Set & Person & Bicycle & Car & Bus & Truck & All \\ 
    \midrule[0.5pt]
    \multirow{4}{*}{Sim-R\textsubscript{test}-by-R (AP@0.3)} & \multirow{2}{*}{RADDet} & 25.13 & 13.69 & \textbf{69.98} & 30.77 & 39.72 & 45.60 \\ 
                       & & 25.13 & \multicolumn{4}{c|}{56.75$^{*}$} & 50.00$^{*}$ \\ \cmidrule(l){2-8} 
                      & \multirow{2}{*}{R. + Sim-R-by-R + Sim-R-by-C} & \textbf{32.03} & \textbf{17.24} & 67.84 & \textbf{46.15} & \textbf{54.70} & \textbf{57.41} \\ 
                      & & \textbf{32.03} & \multicolumn{4}{c|}{\textbf{69.44}$^{*}$} & \textbf{61.29}$^{*}$ \\
    \midrule[0.5pt]
    \multirow{4}{*}{Sim-N\textsubscript{LiDAR}-by-R (AP@0.1)} & \multirow{2}{*}{RADDet} & 17.38 & 0 & 28.45 & 1.18 & 21.21 & 22.11 \\
                       & & 17.38 & \multicolumn{4}{c|}{37.72$^{*}$} & 33.59$^{*}$ \\ \cmidrule(l){2-8} 
                      & \multirow{2}{*}{R. + Sim-R-by-R + Sim-R-by-C} & \textbf{21.56} & 0 & \textbf{36.87} & \textbf{2.36} & \textbf{21.90} & \textbf{26.03} \\ 
                      & & \textbf{21.56} & \multicolumn{4}{c|}{\textbf{43.58}$^{*}$} & \textbf{38.32}$^{*}$ \\
    \midrule[0.5pt]
    \multirow{4}{*}{Sim-N\textsubscript{Camera}-by-R (AP@0.1)} & \multirow{2}{*}{RADDet} & 6.05 & 1.59 & 23.91 & 2.17 & \textbf{25.12} & 18.24 \\
                       & & 6.05 & \multicolumn{4}{c|}{39.29$^{*}$} & 32.53$^{*}$ \\ \cmidrule(l){2-8} 
                      & \multirow{2}{*}{R. + Sim-R-by-R + Sim-R-by-C} & \textbf{7.13} & \textbf{2.38} & \textbf{29.20} & 2.17 & 20.37 & \textbf{20.01} \\ 
                      & & \textbf{7.13} & \multicolumn{4}{c|}{\textbf{41.70}$^{*}$} & \textbf{34.39}$^{*}$ \\
    \bottomrule[1pt]
    \end{tabular}
    \end{adjustbox}
    \caption{3D Object detection AP of RAD boxes on simulated data on RADDet test set scenes and nuScenes~\cite{nuscenes} scenes. Sim-R\textsubscript{test}-by-R: generated on RADDet test set scenes with the radar attribute from $\left \{ \sigma, g, Rs, \lambda \right \} _{R}$. Sim-N-by-R: generated on nuScenes~\cite{nuscenes} v1-mini scenes with the radar attribute from $\left \{ \sigma, g, Rs, \lambda \right \} _{R}$. $^{*}$: treat the above categories as a single category.}
    \vspace{-2mm}
    \label{tab:nuscene}
\end{table*}

\begin{table*}[]
        \centering
    \vspace{2mm}
    \begin{adjustbox}{width=\textwidth}
    \begin{tabular}{c|c|c|cccc|c}
    \toprule[1pt]
    Test Set & Train Set & Person & Bicycle & Car & Bus & Truck & All \\ 
    \midrule[0.5pt]
    \multirow{4}{*}{Sim-R\textsubscript{test}-by-R} & \multirow{2}{*}{RADDet} & 39.09 & 25.86 & 73.19 & \textbf{38.46} & 45.70 & 58.16 \\ 
                       & & 39.09 & \multicolumn{4}{c|}{70.99$^{*}$} & 63.24$^{*}$ \\ \cmidrule(l){2-8} 
                      & \multirow{2}{*}{R. + Sim-R-by-R + Sim-R-by-C} & \textbf{49.34} & \textbf{29.31} & \textbf{84.06} & 30.77 & \textbf{56.10} & \textbf{68.63} \\ 
                      & & \textbf{49.35} & \multicolumn{4}{c|}{\textbf{81.39}$^{*}$} & \textbf{73.90}$^{*}$ \\
    \midrule[0.5pt]
    \multirow{4}{*}{Sim-N\textsubscript{LiDAR}-by-R} & \multirow{2}{*}{RADDet} & 33.10 & \textbf{15.52} & 52.93 & 0.00 & \textbf{72.19} & 40.90 \\
                       & & 33.10 & \multicolumn{4}{c|}{58.51$^{*}$} & 52.14$^{*}$ \\ \cmidrule(l){2-8} 
                      & \multirow{2}{*}{R. + Sim-R-by-R + Sim-R-by-C} & \textbf{42.26} & 8.19 & \textbf{67.90} & \textbf{1.38} & 71.90 & \textbf{49.14} \\ 
                      & & \textbf{42.26} & \multicolumn{4}{c|}{\textbf{67.61}$^{*}$} & \textbf{60.85}$^{*}$ \\
    \midrule[0.5pt]
    \multirow{4}{*}{Sim-N\textsubscript{Camera}-by-R} & \multirow{2}{*}{RADDet} & 13.49 & \textbf{11.81} & 43.85 & 1.09 & \textbf{93.15} & 34.87 \\
                       & & 13.49 & \multicolumn{4}{c|}{62.10$^{*}$} & 48.96$^{*}$ \\ \cmidrule(l){2-8} 
                      & \multirow{2}{*}{R. + Sim-R-by-R + Sim-R-by-C} & \textbf{17.35} & 5.65 & \textbf{55.30} & \textbf{2.17} & 79.44 & \textbf{37.87} \\ 
                      & & \textbf{17.35} & \multicolumn{4}{c|}{\textbf{68.92}$^{*}$} & \textbf{54.36}$^{*}$ \\
    \bottomrule[1pt]
    \end{tabular}
    \end{adjustbox}
    \caption{3D Object detection AP@0.3 of RA boxes on simulated data on RADDet test set scenes and nuScenes~\cite{nuscenes} scenes. Sim-R\textsubscript{test}-by-R: generated on RADDet test set scenes with the radar attribute from $\left \{ \sigma, g, Rs, \lambda \right \} _{R}$. Sim-N-by-R: generated on nuScenes~\cite{nuscenes} v1-mini scenes with the radar attribute from $\left \{ \sigma, g, Rs, \lambda \right \} _{R}$. $^{*}$: treat the above categories as a single category.}
    \vspace{-2mm}
    \label{tab:nuscene2d}
\end{table*}


\end{document}